\documentclass[10pt]{article}   

\usepackage{amsmath}
\usepackage{amssymb}
\usepackage{amsthm}
\usepackage{epsfig}

\newcommand{\prt}{\partial}

\def\L{{\mathcal L}}

\def\ni{{\noindent}}

\numberwithin{equation}{section}
\def\be{\begin{equation}}
\def\ee{\end{equation}}

\usepackage[dvipsnames]{xcolor}

\newcommand{\comment}[1]{}

\newtheorem{theorem}{Theorem}
\newtheorem{prop}[theorem]{Proposition}

\theoremstyle{definition}

\def\XXint#1#2#3{{\setbox0=\hbox{$#1{#2#3}{\int}$}
     \vcenter{\hbox{$#2#3$}}\kern-.5\wd0}}

\begin{document}

\title{{\Large\bf Whitham equations and phase shifts for the Korteweg-deVries equation}}  
\author{Mark~J.~Ablowitz, Justin~T.~Cole, Igor~Rumanov\footnote{corresponding author; e-mail: igor.rumanov@colorado.edu} \\
{\small Department of Applied Mathematics, University of Colorado, Boulder, CO 80309}}
\date{}

%
%
%
%

\maketitle


\begin{abstract}
The semi-classical Korteweg-deVries equation for step-like data is considered with a small parameter in front of the highest derivative. Using perturbation analysis Whitham theory is constructed to higher order. This allows the order one phase and the complete leading order solution to be obtained; the results are confirmed by extensive numerical calculations.

\end{abstract}


\section{Introduction} 
 
Whitham modulation theory has been widely used since it was first developed in 1965 \cite{Whitham65}; see also the classic book~\cite{Whitham74}. An important application of the theory is the Korteweg-de Vries (KdV) equation with small dispersion. Averaging over the fast dynamics that occur over scales on the order of the small dispersion parameter $\epsilon$, Whitham constructed PDEs governing the slowly varying parameters that change over order one space and time scales. 
 These Whitham equations are hyperbolic first order PDEs in space and time. Another breakthrough came several years later when Gurevich and Pitaevskii~\cite{GurPit73} found an important special self-similar solution of Whitham-KdV equations for the step initial condition (IC). The solution is a rarefaction wave solution of Whitham's equations; physically it describes a {\it collisionless} shock wave, also often called a {\it dispersive shock wave} (DSW), which is a consequence of the small dispersion; there is no dissipation. In terms of analysis, at leading order in $\epsilon$ the theory describes a modulated traveling wave KdV solution where the slowly modulated parameters obey the Whitham equations. The fast phase of order $1/\epsilon$ of the traveling wave is determined from these parameters. However, the finite phase shift of order $O(1)$ was not computed, so the description of the leading order solution has remained incomplete until now. 
 
  After this seminal work, Whitham theory developed in many different directions and its applications keep growing. Yet for many years finding the finite phase shifts from Whitham theory was an open problem. In the Whitham approach, this determination requires the computation of higher orders, in particular, the next-to-leading order of Whitham theory which can be viewed as a nonlinear WKB-type expansion. 
 
For {\it integrable} nonlinear PDEs like KdV or e.g.~the one space one time dimensional nonlinear Schr\"odinger equation (1d NLS), there is another approach to find solutions -- the inverse scattering transform (IST), see e.g.~\cite{AbSeg81, AbCl91}. This is equivalent to constructing and solving a Riemann-Hilbert type problem (RHP). It was originally developed for initial value problems with rapidly decaying ICs in both directions in space. Within the RHP framework, the small dispersion limit of KdV with fast decaying ICs was studied by~\cite{DVZ97} extending earlier work, see~\cite{Ven90, LLV93} and references therein, and using the steepest descent approach developed in~\cite{DZ93, DVZ94}. In~\cite{DVZ97}, a trivial, i.e.~constant, finite phase shift was established. The result validity for KdV with decaying ICs was also confirmed by extensive numerics of~\cite{GraKle07}. 

For step-like ICs the long time asymptotics have been considered by IST/RHP methods in~\cite{EGKT13,EGT16}, with a space and time dependent order one phase shift as a result; IST analysis is outside the scope of this paper. Long time and small dispersion are in general different limits. However, the dispersion parameter $\epsilon$ can be removed from KdV equation by rescaling space $x$ and time $t$ variables. Then the step IC is seen as very special since it remains intact by this rescaling. Thus, the Cauchy problem and solution for KdV with step IC depends only on $x/\epsilon$ and $t/\epsilon$. Therefore the long time asymptotics for this IC should apparently be equivalent to the small dispersion limit. In appendix \ref{a:EGT}, we also consider the long time limit result of~\cite{EGT16} for pure step IC and express it in a simpler form. This facilitates a comparison with the result of Whitham theory which we remark upon.

The IST/RHP approach is only applicable to PDEs with known integrable structure. For non-integrable PDEs one has to resort to other methods, and here the nonlinear WKB/Whitham theory approach has been indispensable. Nevertheless, it is important to analyze the well-known PDEs such as KdV and to develop the key ideas in order to pave the way for understanding more complicated models. Moreover, even for the relatively simple situation of KdV with step IC, to our knowledge, the question of finding the $O(1)$ phase shift via Whitham theory still remains unsettled. There is a large number of papers devoted to the leading order theory, associated simulations and experiments involving DSWs, see e.g.~reviews~\cite{ElHoefRev16, ElHSh17} and references therein. However, there are very few dealing with higher-order corrections. An early discussion of higher order effects can be found in~\cite{AbBe70}. Finding the phase shift remains a vital part of the leading order modulated periodic solution.

\par In this paper we derive the higher-order Whitham theory for the KdV equation, the leading order of which was established in \cite{Whitham74, GurPit73}. The key ideas and main results of our approach to higher orders in $\epsilon$ are explained in section 2. Using the (implicit) assumption that the phase shift is included in the total fast phase $\theta\sim1/\epsilon$, we systematically compute the higher order corrections in $\epsilon$ via singular perturbation theory, cf.~e.g.~\cite{AbBe70, Ab2011}. This approach leads to an expansion in powers of $\epsilon^2$ rather than $\epsilon$ {\it for the slow (Whitham) variables}. Then eventually only constant finite (i.e.~$O(1)$, at next-to-leading order in $\epsilon$) phase shift appears in this theory. In other words, every $O(1)$ spacetime varying shift can be absorbed into a redefinition of the (other) basic slow variables and the fast phase $\theta$ determined by them. These results are presented in detail in sections \ref{all1} and \ref{pert}. In the next section \ref{num} we compare the numerical solution of KdV with step IC to the leading order Whitham-GP solution of $O(1/\epsilon)$. The result indicates that the residual phase shift, apart from a constant, is $O(\epsilon)$. If the order unity phase shift for KdV with step IC is constant, then its value can be inferred from the condition at the leading edge of DSW that the solution vanishes; see e.g.~\cite{GurPit73, Grava17} and below in section \ref{num}. 

\par In Whitham theory, the phase shift arises as an integration constant when integrating the leading order ODE eq.~(\ref{eq:u0}) in its fast oscillation phase variable. Any such ``constant" can in general be an arbitrary {\it slow variable} i.e.~a function of space and time which does not change significantly over a period of fast oscillations. Motivated by this observation, in section \ref{all2} we also explore a modified Whitham theory approach by explicitly introducing a spacetime dependent phase shift $\theta_*(x,t)$ into the Whitham-KdV theory from the beginning i.e.~represent the total phase as $\theta = \theta_0/\epsilon + \theta_*$. This changes the look of the higher-order Whitham perturbation theory and leads to the apparent possibility of nontrivial $\theta_*$ dynamics. Such a consideration was initiated back in 1988 by R.~Haberman~\cite{Hab88} who derived equations governing the phase shift $\theta_*$ but did not present any solution. We consider this approach in more detail in section \ref{all2} here but do not find a nonsingular/nontrivial phase shift consistent with our numerical results.

\par As an instructive comparison, we treat in section \ref{lin} the linearized KdV equation in the Whitham framework, see also~\cite{Whitham74}. The linearized KdV equation with e.g.~the same step IC has an exact Fourier solution (the analog of IST solution for nonlinear case). Using Whitham theory, we also derive the corresponding approximate (WKB) solution in the region of fast oscillations. The Whitham approach here yields the exact leading order amplitude and fast phase of the oscillations and allows one to conclude unambiguously that the $O(1)$ phase shift is constant. Only the value of the constant remains undetermined; it can be found from the exact Fourier solution. While the linear problem is simpler, nevertheless it gives additional insights for the nonlinear case. Our conclusions are presented in section 8.

\section{Approach to higher order Whitham theory; main results}

\par We look for an asymptotic solution $u(x,t; \epsilon)$ of the Korteweg-de Vries (KdV) equation,

\be
\prt_tu + 6u\prt_xu + \epsilon^2\prt_{xxx}u = 0,   \label{eq:KdV}
\ee
with fast and slow scales, 
 
\be
u = u(\theta, x, t; \epsilon),   \qquad 0<\epsilon \ll 1,  \label{eq:2.1}
\ee

\ni where the single fast phase $\theta$ is $O(1/\epsilon)$. The fast phase $\theta$ satisfies equations

\be
\prt_x\theta = \frac{k}{\epsilon},  \qquad \prt_t\theta = -\frac{kV}{\epsilon},  \label{eq:2.2}
\ee

\ni with slowly varying quantities $k$ and $V$. Then

\be
\prt_tk+\prt_x(kV) = 0.   \label{eq:k}
\ee
This ``kinematic" equation remains intact at all higher orders in $\epsilon$ since it is just the consequence of definition of the fast phase $\theta$. The other Whitham equations can be derived as secularity conditions ensuring that the solution $u$ is periodic rather than growing in $\theta$. In the next section, by the separation of fast and slow scales method, we find the Whitham equations for KdV to all orders -- see equations (\ref{eq:S2KdV1})-(\ref{eq:PHQd}). 

\par The solution $u$ is expanded in $\epsilon$, $u = u_0 + \epsilon u_1 + \epsilon^2u_2 + \dots$, where $u_0(\theta;x,t;\epsilon)$ is the leading order solution given in eq.~(\ref{eq:u0}), and further corrections satisfy linear equations of the form

$$
\mathcal Lu_n = F_n[u_0, u_1, \dots, u_{n-1}], n = 1, 2, \dots  \qquad  \mathcal L = k^2\frac{d^2}{d\theta^2} + 6u_0(\theta) - V.
$$
The forcing terms $F_n$ at each order depend only on the solution at previous orders of $\epsilon$-perturbation theory. The wavenumber $k$ is determined from eq.~(\ref{eq:k2}) ensuring the constant period of fast oscillations in $\theta$. The solutions to the homogeneous equation $\mathcal Lw=0$ here are known, the first being $w_1=u_0'(\theta)$, and the second, $w_2(\theta)$, is explicitly given in section \ref{pert}, eqs.~(\ref{eq:w2}), (\ref{eq:parw2}). It is of the form

$$
w_2 = K_2u_0'\theta + \phi[u_0],
$$
where $K_2$ is a slow variable and $\phi$ is an explicit non-secular (periodic) function determined by $u_0$. Therefore, using variation of parameters, the particular solutions $u_n$ are given by the integral formulas

\begin{equation}
\label{hot}
u_n = \frac{u_0'(\theta)\int_0^\theta w_2(z)F_n(z)dz - w_2(\theta)\int_0^\theta u_0'(z)F_n(z)dz}{k^2W},
\end{equation}
where $W$ is the Wronskian $W = u_0'w_2' - u_0''w_2$ which is independent of $\theta$. Therefore we can calculate $u = u_0 + \epsilon u_1 + \epsilon^2u_2 + \dots$ at any order in terms of integrals. Moreover, $u_1$ depends on $u_0$; $u_2$ depends on $u_0$ and $u_1$ hence it depends only on $u_0$; similarly $u_j, j=3,4,.... $, depends only on $u_0$. The well-known explicit expression for $u_0$, given by equation (\ref{eq:u0}) is periodic in $\theta$. Then it
can be written as a Fourier series in the variable $\theta$. Hence all successive terms $u_j, j=1,2...$, from equation (\ref{hot}) can be written in terms of Fourier series in $\theta$. We further show in section \ref{pert} and appendix \ref{a:corr1} that we can satisfy the initial/boundary conditions. Thus we have provided a method to solve for $u_n$ and satisfy the initial/boundary conditions  at all orders of $\epsilon$.

\par Then we express the Whitham equations in terms of three Riemann invariants order by order and find the higher order corrections to them given the corrections $u_n$ described above. The nontrivial  corrections at $O(\epsilon^2)$ are given by eq.~(\ref{eq:Wr}) and more explicitly in the next equations after it, which is one of the main results of the paper. 
\par As mentioned above, we first find the Whitham equations to all orders in the original (`physical') variables. These are the conservation of waves eq.~(\ref{eq:k}) and the secularity equations (\ref{eq:S1KdV1}), (\ref{eq:S2KdV1}). Eqs.~(\ref{eq:S1KdV1}) and (\ref{eq:S2KdV1}) themselves are another important result of the paper. This is the first time such equations have appeared in Whitham KdV theory; for a simpler system, the nonlinear Klein-Gordon equation, a similar result was obtained long ago in~\cite{AbBe70}. For KdV, there are only three equations for five dependent variables so they are not closed as they stand. However, these `nonperturbative' Whitham equations can be expanded in the (square of) the small dispersion parameter $\epsilon$ order by order. Their leading order (eqs.~(\ref{eq:s10}), (\ref{eq:s20}) and (\ref{eq:k})) is classical~\cite{Whitham74}. In this case the Whitham equations are closed by using two additional relations, eqs.~(\ref{eq:k2}) and (\ref{eq:Q0}). Eq.~(\ref{eq:k2}) must hold to all orders since it enforces the constant period condition. Eq.~(\ref{eq:Q0}) is replaced at higher orders by corrections found from integrating $u_n$ over the period in $\theta$. This allows one to obtain the corrections to all slow Whitham variables order by order. Eqs.~(\ref{eq:k}), (\ref{eq:S1KdV1}) and (\ref{eq:S2KdV1}) are a convenient means to derive the higher orders of Whitham theory.
\par The fact that these equations contain only $\epsilon^2$, and not $\epsilon$, and the oddness of $u_1$ as a function of $\theta$ lead eventually to the triviality (constancy) of the $O(1)$ phase shift to the order $1/\epsilon$ fast phase of the single DSW solution. This is confirmed by extensive numerics for the special important case of pure step initial condition; the figures and comments in section \ref{numres} are also key results in this paper.

\section{Whitham equations to all orders} \label{all1}


 





After introducing fast and slow scales, the KdV equation takes the form

\be
\left(-\frac{kV}{\epsilon}\prt_{\theta} + \prt_t\right)u + 6u\left(\frac{k}{\epsilon}\prt_{\theta} + \prt_x\right)u + \epsilon^2\left(\frac{k}{\epsilon}\prt_{\theta} + \prt_x\right)^3u = 0.   \label{eq:uKdV1}
\ee

\ni The cubed operator in the third (dispersion) term of eq.~(\ref{eq:uKdV1}) expands in powers of $\epsilon$ as

$$
\epsilon^2\left(\frac{k}{\epsilon}\prt_{\theta} + \prt_x\right)^3 = \frac{k^3}{\epsilon}\prt_{\theta}^3 + 3k^2\prt_{\theta}^2\prt_x + 3k\prt_xk\prt_{\theta}^2 + \epsilon(3k\prt_{\theta}\prt_x^2 + 3\prt_xk\prt_{x\theta} + \prt_{xx}k\prt_{\theta} ) + \epsilon^2\prt_x^3.
$$
We denote $\prt_{\theta}f = f'$ and $\overline f$ the average of $f$ over a period in $\theta$. Eq.~(\ref{eq:uKdV1}) can be written as

\be
k(k^2u'' + 3u^2 - Vu)' + \epsilon kF' = 0,   \label{eq:uF1}
\ee

\ni where

\be
kF' = \prt_tu + 6u\prt_xu + 3k^2\prt_xu''+\frac{3}{2}\prt_x(k^2)u'' + \epsilon\left(3k\prt_{xx}u' + 3\prt_xk\prt_xu' + \prt_{xx}ku'\right) + \epsilon^2\prt_{xxx}u.   \label{eq:F'}
\ee

\ni Imposing periodicity of the solution $u$ and integrating eq.~(\ref{eq:uF1}) over a period in $\theta$ yields an exact (nonperturbative and asymptotic to all orders) secularity condition $\overline{F'} = 0$, or, explicitly,

\be
\prt_t\overline{u} + 3\prt_x(\overline{u^2}) + \epsilon^2\prt_{xxx}\overline{u} = 0.   \label{eq:S1}
\ee

\ni The other needed secularity condition is readily derived when one notices that

\be
u(k^2u'' + 3u^2 - Vu)' = \left(k^2(uu''-(u')^2/2) + 2u^3 - Vu^2/2\right)'   \label{eq:id1}
\ee

\ni is a total derivative in $\theta$. Thus, multiplying eq.~(\ref{eq:uF1}) by $u$ and integrating over the period, one finds the second exact secularity condition $\overline{uF'} = 0$, or, explicitly,  

\be
\prt_t\overline{u^2} + 4\prt_x(\overline{u^3}) - 3\prt_x(k^2\overline{(u')^2}) + 3\epsilon\prt_x[k\overline{(u\prt_xu'-u'\prt_xu)}] + \epsilon^2\prt_x[\prt_{xx}\overline{u^2} - 3\overline{(\prt_xu)^2}] = 0.   \label{eq:S2}
\ee

\ni To derive eq.~(\ref{eq:S2}) we used the identities

$$
\overline{u(3k\prt_{xx} + 3\prt_xk\prt_x + \prt_{xx}k)u'} = \frac{3}{2}(\overline{u(k\prt_{xx} + \prt_xk\prt_x)u'} - \overline{u'(k\prt_{xx} + \prt_xk\prt_x)u}),
$$

$$
k(u\prt_{xx}u'-u'\prt_{xx}u) + \prt_xk(u\prt_{x}u'-u'\prt_{x}u) = \prt_x[k(u\prt_xu'-u'\prt_xu)],
$$

$$
u\prt_{xxx}u = \prt_x[u\prt_{xx}u - (\prt_xu)^2/2] = \prt_x[\frac{\prt_{xx}u^2}{2} - \frac{3(\prt_xu)^2}{2}].
$$
Using the notation $Q = \overline{u}$, $Q_n = \overline{u^n}, n>1$, $G = k^2\overline{(u')^2}$, the two derived secularity equations read:

\be
\prt_tQ + 3\prt_xQ_2 + \epsilon^2\prt_{xxx}Q = 0,   \label{eq:s1}
\ee

\be
\prt_tQ_2 + 4\prt_xQ_3 - 3\prt_xG + 3\epsilon\prt_x\left(k\overline{(u\prt_xu'-u'\prt_xu)}\right) + \epsilon^2\prt_x\left(\prt_{xx}Q_2 - 3\overline{(\prt_xu)^2}\right) = 0.  \label{eq:s2}
\ee
Next we transform the obtained equations to a more convenient form with fewer dependent variables, guided by the well-known leading order KdV manipulations.

\subsection{Transformation of the system (\ref{eq:s1})-(\ref{eq:s2}).}

We integrate eq.~(\ref{eq:uF1}) to get

\be
k^2u'' + 3u^2 - Vu + \epsilon F = C_1,  \label{eq:I1}
\ee

\ni where $C_1=C_1(x,t)$ is an arbitrary integration constant, i.e.~slow variable in our case, and $F$ is defined as a certain antiderivative of eq.~(\ref{eq:F'}),

\be
kF = \prt_tJ_1 + 3\prt_xJ_2 + 3k\prt_x(ku') + \epsilon\left(3k\prt_{xx}u + 3\prt_xk\prt_xu + \prt_{xx}ku\right) + \epsilon^2\prt_{xxx}J_1.   \label{eq:F}
\ee


\ni Functions $J_n$ such that $J_n'=u^n$ contain secular (non-periodic) terms proportional to $\theta$ which we explicitly separate writing

\be
J_n = \overline{u^n}\theta + \hat J_n,   \qquad  \hat J_n' = u^n - \overline{u^n},   \label{eq:hJn}
\ee

\ni so that $\hat J_n$ are periodic. We define all $\hat J_n$ so that  

\be
\overline{\hat J_n} = 0, \ n\in\mathbb N.   \label{eq:avh}
\ee
Next we multiply eq.~(\ref{eq:I1}) by $2u'$ and integrate it again over $\theta$ obtaining 

\be
k^2(u')^2 + 2u^3 - Vu^2 + 2\epsilon(uF - \int uF') = 2C_1u + C_2,   \label{eq:g1}
\ee

\ni where $C_2$ is the arbitrary integration constant (slow variable) and $\int uF'$, as follows from eq.~(\ref{eq:F'}), has the form

$$
k\int uF' = \frac{\prt_tJ_2}{2} + \prt_x(2J_3 - \frac{3}{2}G_1) + 3k^2u\prt_xu' + \frac{3}{2}\prt_x(k^2)uu' +
$$


\be
\epsilon\left( \frac{3}{2}\prt_x(k\int(u\prt_xu'-u'\prt_xu)) + \frac{u}{2}(3k\prt_{xx} + 3\prt_xk\prt_x + \prt_{xx}k)u \right) + \frac{\epsilon^2}{2}\left(\prt_{xxx}J_2 - 3\prt_x\int(\prt_xu)^2\right).   \label{eq:IuF'}
\ee

\ni Here we defined $G_1 = k^2\int(u')^2 = G\theta + \hat G_1$, $\overline{\hat G_1} = 0$, with the previous definitions of $J_n$, now $n=1,2,3$, so that

\be
\overline{J_1} = \frac{Q}{2},  \qquad  \overline{J_2} = \frac{Q_2}{2},  \qquad \overline{J_3} = \frac{Q_3}{2},  \qquad \overline{G_1} = \frac{G}{2}.   \label{eq:Jav}
\ee

\ni We similarly fix the other antiderivatives in eq.~(\ref{eq:IuF'}) as secular $\theta$-term plus periodic ``hat"-term with zero average so that

\be
\overline{\int(u\prt_xu'-u'\prt_xu)} = \frac{\overline{u\prt_xu'-u'\prt_xu}}{2},  \qquad  \overline{\int(\prt_xu)^2} = \frac{\overline{(\prt_xu)^2}}{2}.   \label{eq:Mav}
\ee

\begin{prop}
Secularity conditions eqs.~(\ref{eq:s1}) and (\ref{eq:s2}) for KdV are equivalent to 

\be
\prt_tQ + \prt_x(VQ+C_1) - \epsilon^2\prt_x\left(2\prt_{xx}Q + \frac{3\prt_xk}{k}\prt_xQ + \frac{\prt_{xx}k}{k}Q\right) = 0,   \label{eq:S1KdV1}
\ee

$$
\prt_tP + \prt_xH - \epsilon^2\left[ \prt_tQ_d + \prt_x(VQ_d) + \prt_x\left(\frac{\prt_{xx}P}{2} + \frac{3\prt_xk}{2k}\prt_xP + \frac{\prt_{xx}k}{k}P\right) \right] +
$$

\be
+ \epsilon^4\prt_x\left(\frac{\prt_{xx}Q_d}{2} + \frac{3\prt_xk}{2k}\prt_xQ_d + \frac{\prt_{xx}k}{k}Q_d\right) = 0,  \label{eq:S2KdV1}
\ee
where we denoted 

\be
P=VQ+C_1,  \qquad  H = VP - 3C_2,  \qquad  Q_d = \frac{3\prt_x(k\prt_xQ) + \prt_{xx}kQ}{k}.   \label{eq:PHQd}
\ee
\end{prop}

{\it Proof:}
The average over a period in $\theta$ of eq.~(\ref{eq:I1}) is

\be
3Q_2 - VQ + \epsilon\overline F = C_1.   \label{eq:I1av}
\ee

\ni Integration of eq.~(\ref{eq:g1}) over a period gives relation

\be
G + 2Q_3 - VQ_2 + 2\epsilon(\overline{uF} - \overline{\int uF'}) = 2C_1Q + C_2,   \label{eq:g1av}
\ee

\ni while multiplying eq.~(\ref{eq:I1}) by $u$ and integrating over a period, one gets

\be
-G + 3Q_3 - VQ_2 + \epsilon\overline{uF} = C_1Q.   \label{eq:uI1av}
\ee

\ni Taking into account the secularity condition eq.~(\ref{eq:s1}) in eq.~(\ref{eq:F}) brings the ``forcing" $F$ to the explicitly periodic form,

\be
kF = \prt_t\hat J_1 + 3\prt_x\hat J_2 + 3k\prt_x(ku') + \epsilon\left(3k\prt_{xx}u + 3\prt_xk\prt_xu + \prt_{xx}ku\right) + \epsilon^2\prt_{xxx}\hat J_1.   \label{eq:Fp}
\ee


\ni The last equation integrated over a period becomes very simple,

\be
k\overline F = \epsilon(3k\prt_{xx} + 3\prt_xk\prt_x + \prt_{xx}k)Q.   \label{eq:Fav}
\ee

\ni In turn, taking into account the secularity condition eq.~(\ref{eq:s2}) in eq.~(\ref{eq:IuF'}) lets one bring the quantity $\int uF'$ to explicitly periodic form and its average over a period reads:

\be
k\overline{\int uF'} = \frac{3k^2}{2}\overline{(u\prt_xu'-u'\prt_xu)} + \epsilon\left[\frac{3k}{2}\left(\frac{\prt_{xx}Q_2}{2} - \overline{(\prt_xu)^2}\right) + \frac{3}{2}\prt_xk\frac{\prt_{x}Q_2}{2} + \prt_{xx}k\frac{Q_2}{2}\right].   \label{eq:IuF'av}
\ee

\ni Taking the combination of averaged equations $2\cdot(\ref{eq:uI1av}) - (\ref{eq:g1av})$ yields

\be
4Q_3 - 3G - VQ_2 + 2\epsilon\overline{\int uF'} = -C_2.   \label{eq:*2}
\ee
Upon using eq.~(\ref{eq:IuF'av}), eq.~(\ref{eq:*2}) acquires the form containing exactly the combination entering the secularity equation (\ref{eq:s2}),

\be
4Q_3 - 3G + 3\epsilon k\overline{(u\prt_xu'-u'\prt_xu)} - 3\epsilon^2\overline{(\prt_xu)^2} = VQ_2-C_2 - \frac{\epsilon^2}{k}\left(\frac{3}{2}\prt_x(k\prt_xQ_2) + \prt_{xx}kQ_2 \right).   \label{eq:*22}
\ee
We substitute the right-hand side of eq.~(\ref{eq:*22}) into eq.~(\ref{eq:s2}) and the last becomes

\be
\prt_tQ_2 + \prt_x(VQ_2-C_2) - \epsilon^2\prt_x\left(\frac{\prt_{xx}Q_2}{2} + \frac{3\prt_xk}{2k}\prt_xQ_2 + \frac{\prt_{xx}k}{k}Q_2\right) = 0.   \label{eq:s22}
\ee

\ni Finally we use eqs.~(\ref{eq:I1av}) and (\ref{eq:Fav}) to express $Q_2$ as

\be
Q_2 = \frac{VQ+C_1}{3} - \epsilon^2\left( \frac{\prt_x(k\prt_xQ)}{k} + \frac{\prt_{xx}k}{3k}Q \right)   \label{eq:Q2}
\ee
and substitute it into eqs.~(\ref{eq:s1}) and (\ref{eq:s22}). Thus, we obtain the secularity conditions in their final form of eqs.~(\ref{eq:S1KdV1}) and (\ref{eq:S2KdV1}), as claimed.

\bigskip

\par The secularity equations (\ref{eq:S1KdV1}), (\ref{eq:S2KdV1}) and the kinematic equation (\ref{eq:k}) comprise exact nonperturbative Whitham-KdV equations in physical variables $V, C_1, C_2, k$ and $Q$. However, the system of Whitham PDEs is not closed as it stands. Still it is a very convenient starting point to get the Whitham equations to any needed higher order in $\epsilon$. We see that the system (\ref{eq:S1KdV1}), (\ref{eq:S2KdV1}) is perturbed only by $\epsilon^2$. This suggests that under a perturbation expansion of $u$ in powers of $\epsilon$ the secularity conditions will have all nontrivial higher-order corrections expanded in $\epsilon^2$. We will demonstrate in some detail that this is indeed the case in the next section. This, in particular, will imply that the first nontrivial correction to the fast phase $\theta$ is going to be of order $\epsilon$. As for an order $O(1)$ phase shift, it can then only be a pure constant, the value of which only affects the initial/boundary conditions but not the equations. As we will see in section \ref{numres}, this picture is consistent with the numerical results.

\section{Perturbations of Whitham variables} \label{pert}

Let $u = u_0 + \epsilon \tilde u$, where $u_0$ is {\it defined} as the solution of first order ODE

\be
k^2(u_0')^2 = - 2u_0^3 + Vu_0^2 + 2C_1u_0 + C_2,   \label{eq:g10}
\ee
compare with eq.~(\ref{eq:g1}). Equation (\ref{eq:g10}) includes all leading order terms of eq.~(\ref{eq:g1}), therefore $u-u_0$ starts at order $\epsilon$ indeed. Let $\lambda_1\le\lambda_2\le\lambda_3$ be the roots of the cubic in the right hand side of eq.~(\ref{eq:g10}). They are related to $V$, $C_1$ and $C_2$ as 


{\small \be
\frac{V}{2} = e_1 \equiv \lambda_1 + \lambda_2 + \lambda_3,  \quad  C_1=-e_2 \equiv -(\lambda_1\lambda_2+\lambda_2\lambda_3+\lambda_3\lambda_1),  \quad \frac{C_2}{2}=e_3\equiv \lambda_1\lambda_2\lambda_3.  \label{eq:e123}
\ee}

\ni The normalization of the elliptic cnoidal solution to eq.~(\ref{eq:g10}) with unit period in $\theta$ implies that
\be
k^2 = \frac{\lambda_3-\lambda_1}{8K^2(m)},   \label{eq:k2}
\ee
where $K(m)$ is the first complete elliptic integral; hence the solution to eq.~(\ref{eq:g10}) is 
\be
u_0 = \lambda_2 + (\lambda_3-\lambda_2)\text{cn}^2(2K(m)\theta; m),  \qquad m = \frac{\lambda_3-\lambda_2}{\lambda_2-\lambda_1}.   \label{eq:u0}
\ee
Also we have $Q=Q_0+\epsilon \tilde Q$, where $Q_0 = \overline{u_0}$ so that

\be
Q_0 = \lambda_1 + (\lambda_3-\lambda_1)\frac{E(m)}{K(m)}   \label{eq:Q0}
\ee
in terms of the first and second complete elliptic integrals $K(m)$ and $E(m)$. Then, keeping only terms starting at leading order in eqs.~(\ref{eq:S1KdV1}), (\ref{eq:S2KdV1}) and adding the relations (\ref{eq:k2}), (\ref{eq:Q0}), one gets the closed familiar Whitham-KdV system consisting of eqs.~(\ref{eq:k}), (\ref{eq:k2}), (\ref{eq:Q0}), and equations

\be
\prt_tQ_0 + \prt_x(VQ_0+C_1) = 0,   \label{eq:s10}
\ee

\be
\prt_t(VQ_0+C_1) + \prt_x\left(V(VQ_0+C_1)-3C_2\right) = 0.   \label{eq:s20}
\ee
At this point the Whitham-KdV system can be viewed as three equations (\ref{eq:k}), (\ref{eq:s10}) and (\ref{eq:s20}) for three unknowns $\lambda_j, j=1,2,3$, where the constraints eqs.~(\ref{eq:k2}) and (\ref{eq:Q0}) were substituted. We introduce the KdV Riemann invariants $r_1\le r_2\le r_3$ \cite{Whitham74, GurPit73} such that in terms of them this system diagonalizes with respect to space and time derivatives and takes form

$$
\prt_tr_j + v_j(r_1,r_2,r_3)\prt_xr_j = 0.
$$
Explicitly $r_j$-variables are linearly related with the cubic roots $\lambda_i$,
\be
\lambda_1 = r_1+r_2-r_3,  \qquad \lambda_2 = r_1-r_2+r_3,  \qquad \lambda_3 = -r_1+r_2+r_3,  \label{eq:rKdV}
\ee
and the other involved slow variables are the following functions of the three $r$-s:

$$
m = \frac{r_2-r_1}{r_3-r_1}, \qquad k^2 = \frac{r_3-r_1}{4K^2(m)},   \qquad Q_0 = r_2 - (r_3-r_1) + 2(r_3-r_1)\frac{E(m)}{K(m)}, 
$$

$$
V = 2(r_1+r_2+r_3),  \qquad C_1 = r_1^2+r_2^2+r_3^2-2(r_1r_2+r_2r_3+r_3r_1),
$$

\be
C_2 = 2(-r_1^3-r_2^3-r_3^3+r_1^2(r_2+r_3)+r_2^2(r_3+r_1)+r_3^2(r_1+r_2)-2r_1r_2r_3).   \label{eq:ph-r}
\ee

On the other hand, the all orders system of Whitham equations obtained in the previous section is

\be
\prt_tk+\prt_x(kV) = 0,   \label{eq:kV}
\ee

\be
\prt_tQ + \prt_x(VQ+C_1) - \epsilon^2\Phi_1 = 0,    \label{eq:S1KdV2}    
\ee

\be
\prt_t(VQ+C_1) + \prt_x[V(VQ+C_1) - 3C_2] - \epsilon^2\Phi_2 = 0,  \label{eq:S2KdV2} 
\ee
where $\Phi_1$ and $\Phi_2$ are also explicit in terms of $k, V, C_1$ and $Q$,

\be
\Phi_1 = \prt_x\left(2\prt_{xx}Q + \frac{3\prt_xk}{k}\prt_xQ + \frac{\prt_{xx}k}{k}Q\right),   \label{eq:Phi1}
\ee


{\small\be
\Phi_2 = \prt_tQ_d + \prt_x\left(VQ_d + \frac{\prt_{xx}P}{2} + \frac{3\prt_xk}{2k}\prt_xP + \frac{\prt_{xx}k}{k}P\right) - \epsilon^2\prt_x\left(\frac{(k\prt_{xx} + 3\prt_xk\prt_x + 2\prt_{xx}k)Q_d}{2k}\right),   \label{eq:Phi2} 
\ee}

$$
P\equiv VQ+C_1,   \qquad  Q_d \equiv \frac{3\prt_x(k\prt_xQ) + \prt_{xx}kQ}{k}.   
$$
It is remarkable that only the mean of the KdV solution $Q\equiv \overline u$ here has to be found using the higher order corrections to KdV itself, no other information from them is needed. These corrections $u_n$ to $u = u_0 + \epsilon u_1 + \epsilon^2u_2 + \dots$ are found from

\be
k^2u_1'' + (6u_0-V)u_1 + F_1 = 0,   \label{eq:o1}
\ee

\be
k^2u_2'' + (6u_0-V)u_2 + F_2 = 0,   \label{eq:o2}
\ee
and so on; the forcing terms $F_1$ and $F_2$ are

\be
kF_1 = \prt_t\hat J_1^0 + 3\prt_x\hat J_2^0 + 3k^2\prt_xu_0' + 3k\prt_xku_0',    \label{eq:f1}
\ee

\be
kF_2 = 3u_1^2 + \prt_t\hat J_1^1 + 3\prt_x\hat J_2^1 + 3k^2\prt_xu_1' + 3k\prt_xku_1' + 3k\prt_{xx}u_0 + 3\prt_xk\prt_xu_0 + \prt_{xx}ku_0,    \label{eq:f2}
\ee

$$
(\hat J_1^n)' = \hat u_n = u_n - \overline u_n,  \qquad  (\hat J_2^n)' = \hat {u_n^2} = u_n^2 - \overline {u_n^2},  \qquad \overline{\hat J_m^n} = 0.
$$
Using variation of parameters, the solution to eqs.~(\ref{eq:o1}), (\ref{eq:o2}) is given by the integral formulas

\be
u_1 = \frac{u_0'(\theta)\int_0^\theta w_2(z)F_1(z)dz - w_2(\theta)\int_0^\theta u_0'(z)F_1(z)dz}{k^2W},    \label{eq:u1s}
\ee

\be
u_2 = \frac{u_0'(\theta)\int_0^\theta w_2(z)F_2(z)dz - w_2(\theta)\int_0^\theta u_0'(z)F_2(z)dz}{k^2W},    \label{eq:u2s}
\ee
and so on, where $u_1$ depends on $u_0$ only, $u_2$ depends on $u_0$ and $u_1$ only and every subsequent $u_n$ depends only on all the previous ones, $u_0$ through $u_{n-1}$. In the last two formulas, $w_2(\theta)$ is the second homogeneous solution of the linear operator $\L=k^2d^2/d\theta^2 + 6u_0 - V$ in eqs.~(\ref{eq:o1}), (\ref{eq:o2}) etc., the first being $u_0'(\theta)$. Explicitly, $w_2$ is given by

\be
w_2 = \alpha (u_0'\hat J_1^0 - 2u_0^2) + (\alpha Q_0+\beta)u_0'\theta + \delta u_0 + \chi,   \label{eq:w2}
\ee
it can be normalized so that the constants in eq.~(\ref{eq:w2}) are


\be
\alpha = \frac{V^2}{3}+4C_1,   \quad  \beta = \frac{VC_1}{3}+3C_2,  \quad  \delta = 2\beta + V\alpha,  \quad \chi = \frac{4C_1\alpha - V\beta}{3}. 
\label{eq:parw2}  
\ee
This $w_2$ can be obtained e.g.~by looking for a solution of a form like eq.~(\ref{eq:w2}), which gives $\delta, \chi$ in terms of $\alpha,\beta$ as in eq.~(\ref{eq:parw2}) and fixes the ratio $\alpha/\beta$ as $(VC_1+9C_2)\alpha=(V^2+12C_1)\beta$. Then the Wronskian $W = u_0'w_2' - u_0''w_2$, which is a slow variable (independent of $\theta$), is explicitly given by (when $\alpha$ and $\beta$ are chosen as in eq.~(\ref{eq:parw2}))




\be
k^2W = (VC_2-4C_1^2/3)\alpha + (VC_1/3+3C_2)\beta.   \label{eq:Wronsk}
\ee
The lower limits of integration in eqs.~(\ref{eq:u1s}) and (\ref{eq:u2s}) ensure that $u_n(x,0)=0$ for $n\ge1$ i.e.~the IC has to be satisfied by the leading order solution $u_0$. The above and the given formulas for $u_1$, $u_2$ etc.~are true when $\theta(x,0)=0$, see the next section and eq.~(\ref{eq:phase0}) in particular, when the fast phase is ``born'' after wave breaking at time $t=0$, which is the situation we study here. More general IC consideration is given in appendix \ref{a:corr1}. There it is also shown that $u_1$, $u_2$, $\dots$ given by eqs.~(\ref{eq:u1s}), (\ref{eq:u2s}) etc.~are indeed periodic hence non-secular.

\par Since $u_1$ is odd in $\theta$ because $F_1$ is, we have $\overline u_1=0$. This is also the case for all further odd order corrections $u_{2n+1}$. Therefore the total mean $Q$ can be written as 

\be
Q = \overline u_0 + \epsilon^2q = Q_0 + \epsilon^2(q_2 + \epsilon^2q_4 + O(\epsilon^4)),   \qquad  q_2 = \overline u_2, \quad q_4 = \overline u_4, \quad \dots   \label{eq:Qq}
\ee
where $q$ has to be determined order by order by solving the higher order corrections to the KdV equation or rather to its first integral eq.~(\ref{eq:I1}); i.e.~the function $q$ is determined by averages of even $u$-corrections $u_{2n}$.
\par The Whitham equations can be represented order by order in terms of the three Riemann variables $r_j$, $j=1,2,3$, defined exactly as in the leading order equations above. We transform eqs.~(\ref{eq:kV}), (\ref{eq:S1KdV2}) and (\ref{eq:S2KdV2}) using eqs.~(\ref{eq:rKdV}) and (\ref{eq:ph-r}) and obtain the Whitham equations diagonalized to leading order in Riemann variables:

\be
\prt_tr_j + v_j\prt_xr_j + \frac{\epsilon^2(X_2 + 6(r_j-r_l-r_m)X_1)}{24(r_j-r_l)(r_j-r_m)\prt_jk/k} = 0,   \qquad j\neq l \neq m \neq j, \quad j=1,2,3 \label{eq:Wr}
\ee

$$
X_1 \equiv \prt_tq + \prt_x(Vq) - \Phi_1,  \qquad  X_2 \equiv \prt_t(Vq) + \prt_x(V^2q) - \Phi_2,   \qquad  \prt_jk/k \equiv \frac{\prt k}{k\prt r_j}. 
$$
The quantity $q$ defined in eq.~(\ref{eq:Qq}) is also expressed order by order in terms of the $r_j$-variables by solving eqs.~(\ref{eq:o1}), (\ref{eq:o2}) and so on.

Asymptotically, the Whitham equations, e.g.~in Riemann form eq.~(\ref{eq:Wr}), can be solved order by order in the $r$-variables. The first corrections come at order $\epsilon^2$ which implies a nontrivial (nonconstant) correction to the fast phase $\theta$ being of order $\epsilon$ only. Thus, the most important $O(1)$ phase shift has to be a pure constant rather than a slow variable by this consideration. The value of the constant is then fixed by initial/boundary conditions. E.g.~in the case of pure step ICs of~\cite{GurPit73} or section \ref{num}, the constant is fixed by the condition at the leading (solitonic) edge of the DSW. We also note that when solving order by order, $u_n$ do not have any singularities for $0<m<1$. E.g.~Whitham equations to order $O(\epsilon^2)$ can be presented in the form of eq.~(\ref{eq:Wr}) where


\small{$$
X_1 = \prt_tq_2 + \prt_x(Vq_2) - \prt_x\left(2\prt_{xx}Q_0 + \frac{3\prt_xk}{k}\prt_xQ_0 + \frac{\prt_{xx}k}{k}Q_0\right),   \qquad  q_2 = \overline{u_2},  \qquad  P_0=VQ_0+C_1, 
$$}

\small{$$
X_2 = \prt_t(Vq_2) + \prt_x(V^2q_2) - \prt_tQ_d^0 + \prt_x\left(VQ_d^0 + \left[\frac{\prt_{xx}}{2} + \frac{3\prt_xk}{2k}\prt_x + \frac{\prt_{xx}k}{k}\right]P_0\right),    \quad  Q_d^0 = \frac{3\prt_x(k\prt_xQ_0) + \prt_{xx}kQ_0}{k} 
$$}
and $k, Q_0, V, C_1$ are given in terms of $r_j, j=1,2,3$, by eq.~(\ref{eq:ph-r}). One could obtain the Fourier series for $u_2$ and take the average in $\theta$ to get $q_2$, leaving only slow variables.
\par The system (\ref{eq:Wr}) can, in principle, also be solved numerically. In this regard we consider an iteration where the terms without $\epsilon^2$ are iterates at level $n+1$ and the terms with $\epsilon^2$ are iterates at level $n$. At $n=0$ we take the perturbing iterate with $\epsilon=0$; i.e.~we have our unperturbed solution. The $n=1$ term is solved by calculating the perturbed terms with
\[u=u_0+\epsilon u_1+\epsilon^2 u_2,   \qquad   Q- \overline{u_0} =\epsilon^2 q\]
with ODEs and definitions given above for eq.~(\ref{eq:Wr}), $u_1,u_2$ etc. 

\section{Step initial value problem. Numerical results}  \label{num}

Consider the step IC for the KdV equation,

\be
u(x,0) = \left\{ \begin{array}{cc} 1, & x<0 \\ 0, & x>0.  \end{array} \right.   \label{eq:stepIC}
\ee

\ni Then Whitham theory gives the famous Gurevich-Pitaevskii (GP) DSW solution involving the leading order cnoidal function with modulated parameters. In the GP solution, $r_1=0$, $r_3=1$ and $r_2=m=m(x/t)$ is implicitly given by

\be
\frac{x}{t} = v_2(m) = 2(1+m) - \frac{4m(1-m)K(m)}{E(m)-(1-m)K(m)}.   \label{eq:GP}
\ee

\ni Only the leading order phase $\theta_0(x,t)/\epsilon$ is known in the GP solution. It can be determined from $k$ and $V$ by formula

\be
\frac{\theta_0(x,t)}{\epsilon} = \int_0^x\frac{k(\eta, t)}{\epsilon}d\eta - \int_0^t\frac{k(0,\tau)V(0,\tau)}{\epsilon}d\tau.   \label{eq:phase}
\ee
The lower limits of integration here reflect the fact that the fast phase is ``born" at time $t=0$ at the jump point $x=0$. Explicitly

\be
\frac{\theta_0(x,t)}{\epsilon} = -\left(r_3-r_1\right)^{3/2}\frac{t}{\epsilon K'(m)} = -\frac{t}{\epsilon}\cdot\frac{2m(1-m)}{E(m) - (1-m)K(m)}.   \label{eq:phase0}
\ee
Hence $\theta(x,0)=0$ and the lower limits of integration in eqs.~(\ref{eq:u1s}), (\ref{eq:u2s}) vanish.
Considering the leading order modulated traveling wave solution $u_0$ in eq.~(\ref{eq:u0}) which now takes the form

\be
u_0 = 1 - m + 2m~ {\rm cn}^2\left( 2 K(m) \theta; m \right),   \label{eq:u0m}
\ee
we observe that it satisfies the step IC behind the step for $x<0$ where the amplitude of the cnoidal function $\lambda_3-\lambda_2 = 2(r_2-r_1) = 0$. In order for $u_0$ to satisfy also the IC $u_0=0$ in front of the step for $x>0$, where $m \to 1$, one needs to have $\lim_{m \to 1} \text{cn}(2K(m)\theta; m)=0$ there which implies $\theta(x,0) = \pm1/2$ for $x>0$. Thus, for consistency with eq.~(\ref{eq:phase0}) for the fast phase, the total phase $\theta$ must contain a constant order one phase shift $\pm1/2$ i.e.~$\theta(x,t) = \theta_0(x,t)/\epsilon \pm 1/2 + \dots$, where dots stand for any possible further corrections of order zero or higher in $\epsilon$ which must vanish identically at $t=0$. Then all higher order corrections to the leading order solution $u_0$ must also vanish identically at $t=0$.

\par The numerics presented below for the step IC clearly demonstrate that the additional phase shift described by dots in the last equation is of order $\epsilon$. We numerically determine the positions of the maxima of the computed solution $u(x,t)$ and compare them with the maxima obtained for the approximate theoretical leading order solution $u_0$ with the phase $\theta$ given by $\theta=\theta_0/\epsilon + 1/2$. 

\subsection{Numerical methods}

\begin{figure} [ht]
\centering
\includegraphics[scale=.2]{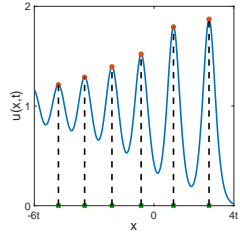}
\caption{Locations of DSW maxima, $x_{\max}$. The maxima in terms of the modulus are computed via eq.~(\ref{convert_x_to_m}).}
\label{DSW_diagram}
\end{figure}

This section describes the numerical scheme used to solve the KdV equation eq.~(\ref{eq:KdV}) with step initial data eq.~(\ref{eq:stepIC}).
The idea is to introduce a source function that ``cancels out'' the solutions of eq.~(\ref{eq:KdV}) at $x \to \pm \infty$. This ``cancellation'' function reformulates the problem into one that has zero boundary conditions on both sides. From there we utilize fast Fourier methods to approximate spatial derivatives and implement an exponential time-differencing Runge-Kutta (ETDRK) scheme to integrate. The ETD class of methods are an ideal for problems like KdV since they solve the rapidly oscillating part of the equation exactly.

\begin{figure} [ht]
\centering
\includegraphics[scale=.25]{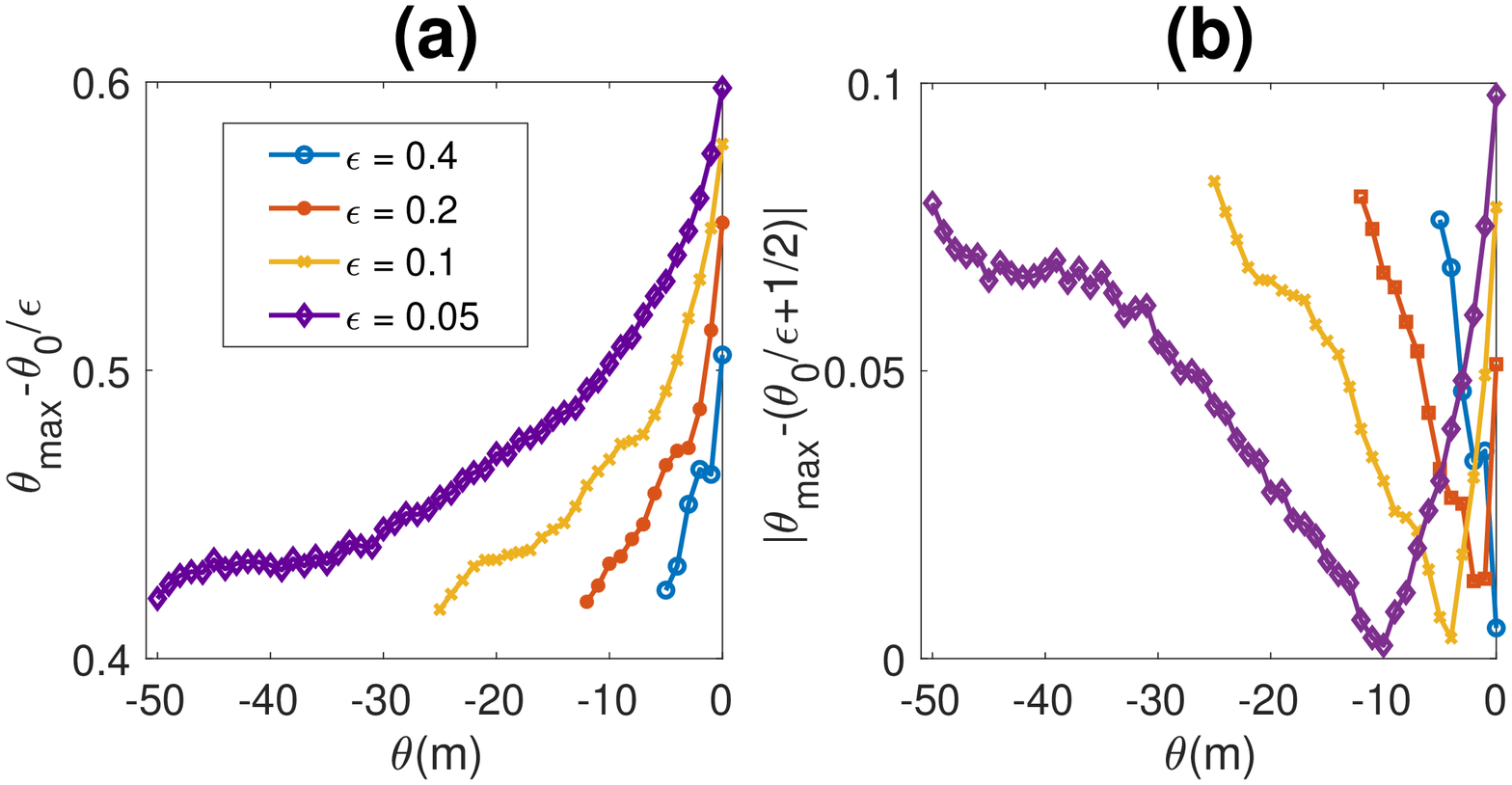}    \qquad  \includegraphics[scale=.25]{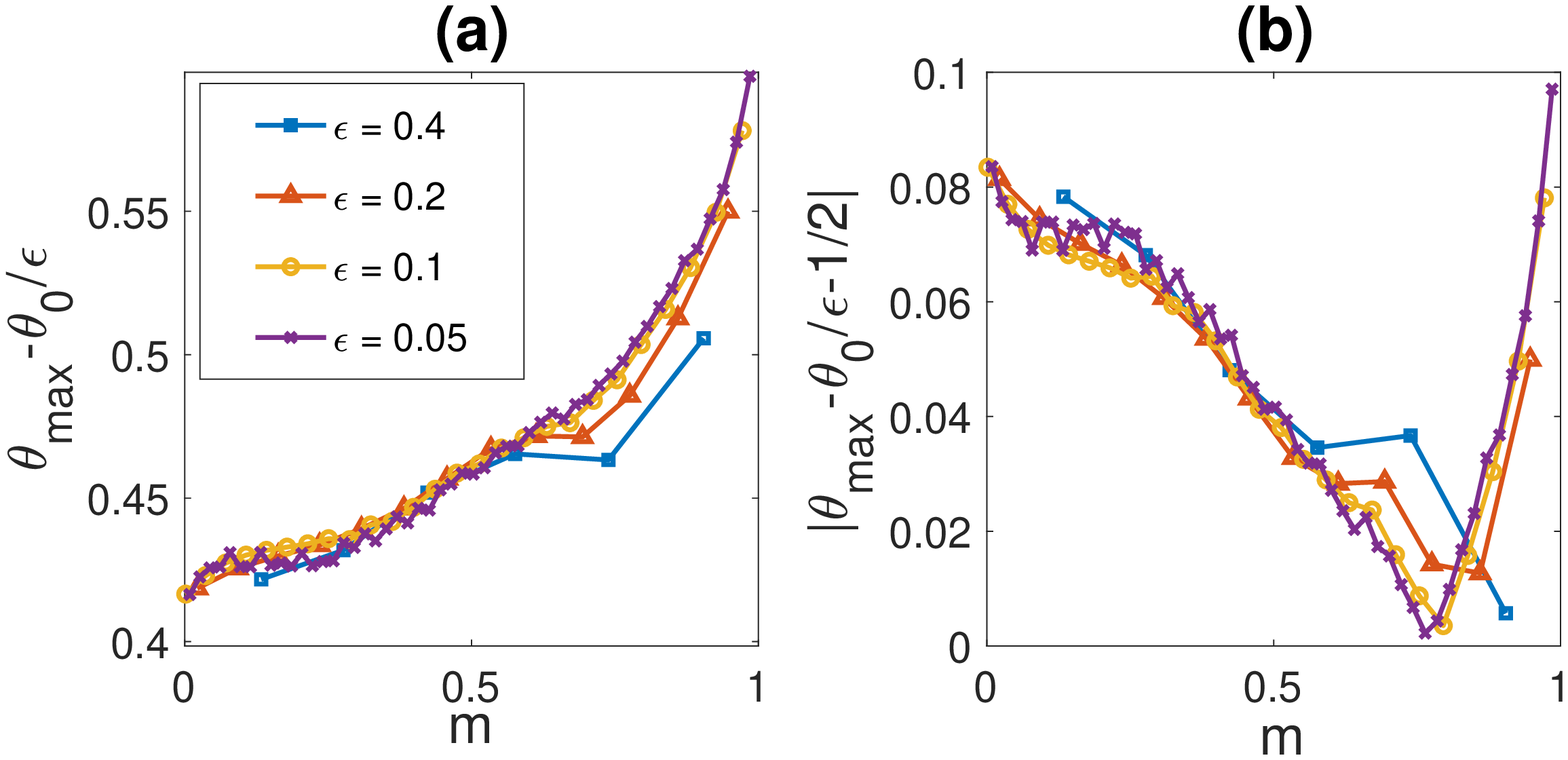}
\caption{{\bf Left:} The difference between the numerically computed DSW maxima (integers) and the asymptotic approximations for different values of $\epsilon$ at time $t = 1$: (a) $\theta_0/\epsilon$ and (b) $\theta_0/\epsilon + 1/2$ vs.~the function $\theta_0(m)$ given by eq.~(\ref{eq:phase0}). This is evaluated at the numerically computed maxima locations shown in Fig.~\ref{DSW_diagram}. {\bf Right:} The difference between the numerically computed DSW maxima (integers) and the asymptotic approximations for different values of $\epsilon$ at time $t = 1$: (a) $\theta_0/\epsilon$ and (b) $\theta_0/\epsilon + 1/2$ vs.~the modulus $m$. }
\label{combined_phase_shift}
\end{figure}
%


To begin, consider decomposing the solution $u(x,t)$ as 
\begin{equation}
\label{u_expand}
u(x,t) = v(x,t) + w(x) ,
\end{equation}
with the accompanying boundary conditions: $v \rightarrow 0$ as $|x| \rightarrow \infty$, and
\begin{equation}
w(x) \rightarrow \begin{cases}
 1 & x \rightarrow - \infty \\
 0 &x  \rightarrow  + \infty
\end{cases} .
\end{equation}
The localized function $v(x,t)$ is unknown and must be solved for. The ``cancellation'' function $w(x)$ is chosen with the appropriate boundary conditions. We typically take something simple that is easy to differentiate exactly, such as 
\begin{equation}
w(x) = \frac{1 - \tanh(x)}{2} .
\end{equation}
For solutions of the form in eq.~(\ref{u_expand}), the governing equation (\ref{eq:KdV}) is expressed as
\begin{equation}
\label{reformulate_kdv}
v_t + 3 ( v^2 )_x + 6 (v w)_x + \epsilon^2 v_{xxx} = -  3 (w^2)_x - \epsilon^2  w_{xxx} ,
\end{equation}
with initial condition: $v(x,0) = u(x,0) - w(x)$. The initial step $u(x,0)$ is numerically approximated by a sharp (relative to $\epsilon$) hyperbolic tangent function of 
\begin{equation}
\label{num_IC_u}
u(x,0) = \frac{1 - \tanh(x/\delta)}{2} ,
\end{equation}
where $\delta = \epsilon/10$. Note that this equation has zero boundary conditions at infinity. We approximate all spatial derivatives of $v(x,t)$ by Fourier methods. A wide computational domain is used to ensure waves from the linear edge of the DSW do not propagate through the periodic boundary conditions and back into the DSW region. We integrate (\ref{reformulate_kdv}) by the ETDRK4 scheme described in \cite{trefethen}. 

After numerically solving for $u(x,t)$, the maxima values in the DSW region are computed. These points are all the local maxima located within the interval $- 6t \le x \le 4 t$. A diagram illustrating this is shown in Fig.~\ref{DSW_diagram}. A maximum value $x_{\max}$ at time $t$ is converted to corresponding value of the elliptic modulus $m_{\max}$ through the GP formula eq.~(\ref{eq:GP})
\begin{equation}
\label{convert_x_to_m}
\frac{x_{\max}}{t} = 2(1 + m_{\max}) + \frac{4 m_{\max} (1 - m_{\max})}{1 - m_{\max} - E(m_{\max})/K(m_{\max})},
\end{equation}
using a root-finding method to invert eq.~(\ref{convert_x_to_m}). The maxima of the asymptotic solution eq.~(\ref{eq:u0m}) occur at integer values of $\theta$, i.e.~where
\begin{equation}
\theta_{\max} = \theta( m_{\max}) = \frac{\theta_0}{\epsilon} + \theta_* + \cdots \bigg|_{m = m_{\max}}  = n \in \mathbb{Z} .
\end{equation}
To approximate the phase shift $\theta_*$ we take the difference between a set of integers and $\theta_0 / \epsilon$. It is arbitrary where to begin the integers; different starting values result in $\theta_*$ being shifted by an integer amount. We take $n = 0$ at the largest DSW peak, that is the maximum nearest to $m = 1$. The integers decrease as $m$ decreases, resulting in the phase values $\theta_{\max}  = \{ 0 , -1 , -2 , \dots \}.$

\begin{figure} [ht]
\centering
\includegraphics[scale=.25]{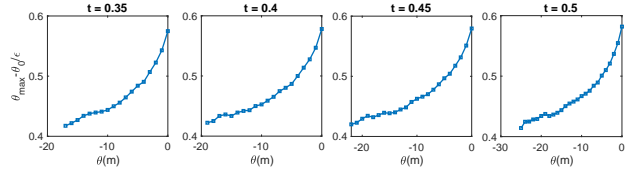}
\caption{Difference between DSW maxima $\theta_{\max}$ and $\theta_0/\epsilon$ at different times for $\epsilon = 0.05$. }
\label{curve_shapshots_eps_005}
\end{figure}

\begin{figure} [ht]
\centering
\includegraphics[scale=.25]{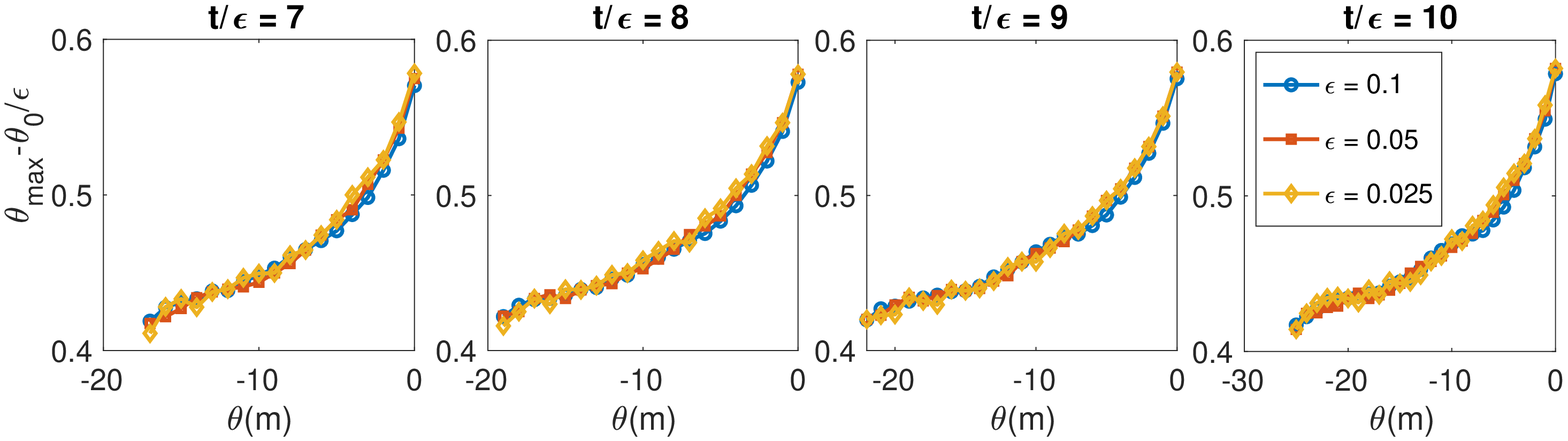}
\caption{Difference between $\theta_{\max}$ and $\theta_0/\epsilon$ for different values of $t / \epsilon$. Each curve corresponds to a different value of $\epsilon$, and $t$ is chosen so that $t/\epsilon$ is constant.}
\label{curve_snapshots_eps01_005}
\end{figure}

\subsection{Numerical Results}     \label{numres}

The difference in the maximum phase values, $\theta_{\max}$, and $\theta_0/ \epsilon$ is shown in Fig.~\ref{combined_phase_shift} Left (a) at time $t = 1$. As $\epsilon$ decreases, the number of maxima points increases. For each value of $\epsilon$, the value of $\theta_*$ is found to be nonzero and approximately $1/2$, i.e. $\theta_0/\epsilon$ takes half-integer values at the maxima. A comparison between $\theta_{\max}$ and the approximation $\theta_0/ \epsilon + 1/2$ is given in Fig.~\ref{combined_phase_shift} Left (b). Overall there is excellent agreement between the two curves, with error less than $0.1$ (i.e.~$\leq 2 \epsilon$) for all cases considered. The largest disagreement comes near $m =0$, where there is no improvement as $\epsilon$ decreases, and $m = 1$, where there is a slight growth in the error as $\epsilon$ decreases. Another enlightening view of these results is presented in Fig.~\ref{combined_phase_shift} Right. Here the abscissa is converted to values of $m$ using the GP formula (\ref{eq:phase0}). We see that the results are tending to an asymptotic result as $\epsilon \to 0$. There might be intermediate/transition regions near $m=1$ and perhaps $m=0$ where the solution is governed by different scalings and the formulae in the DSW region may not be uniformly applicable.

\par Next we seek to establish that, for large times, $\theta_*$ depends only on $m$, and not {\it both} $m$ and $t$. The difference between $\theta_{\max}$ and $\theta_0/ \epsilon$ at several different times is shown in Fig.~\ref{curve_shapshots_eps_005}. The snapshot series in Fig.~\ref{curve_shapshots_eps_005} indicates that once $t / \epsilon$ is sufficiently large, the corrections $\theta_* + O(\epsilon)$ approach a steady state. The small deviations can be attributed to the sensitive nature of tracking the maximum point $x_{\max}$ and converting it into a modulus $m_{\max}$.

\par Finally, numerics indicate that the mode profile has time dependence $\theta_*(m (t/\epsilon))$. In Fig.~\ref{curve_snapshots_eps01_005} the phase difference is shown for the same value of $t / \epsilon$, but different values of $\epsilon$. What these figures demonstrate is that the phase does not depend on $t$ and $\epsilon$ separately.

\section{Whitham equations with additional phase shift}  \label{all2}

There is a variation of Whitham theory which explicitly introduces a nontrivial phase shift that is independent of the leading order fast phase. This was considered e.g.~by Haberman in~\cite{Hab88}. We employ this idea here. This approach is a natural generalization of the previous considerations: when one integrates the leading order ODE eq.~(\ref{eq:g10}), the ``integration constant" $\theta_*$ is in general a slow function of $x, t$, independent of $\theta$. Below we consider this more general theory.
\par We start again with eq.~(\ref{eq:KdV}) and now look for a solution $u$ with fast and slow scales in the form

\be
u = u(\theta, x, t; \epsilon),   \qquad \theta = \frac{\theta_0}{\epsilon} + \theta_*,   \qquad 0<\epsilon \ll 1,  \label{eq:2.1v}
\ee

\ni where we define 

\be
\prt_x\theta_0 = k,  \qquad \prt_t\theta_0 = -kV,  \label{eq:2.2v}
\ee

\ni and $k$ and $V$ are slowly varying quantities. Then the consideration parallels that of section \ref{all1}. Again we have eq.~(\ref{eq:k}) and the other Whitham equations are derived as secularity conditions ensuring that the solution $u$ is periodic rather than growing in $\theta$.
\par The KdV equation now, after introducing fast and slow scales, takes the form

\be
\left((-\frac{kV}{\epsilon}+\prt_t\theta_*)\prt_{\theta} + \prt_t\right)u + 6u\left((\frac{k}{\epsilon}+\prt_x\theta_*)\prt_{\theta} + \prt_x\right)u + \epsilon^2\left((\frac{k}{\epsilon}+\prt_x\theta_*)\prt_{\theta} + \prt_x\right)^3u = 0.   \label{eq:uKdV}
\ee
We will again denote $\prt_{\theta}f = f'$ and will also use the notation 
$$
\tilde k = k + \epsilon\prt_x\theta_*.   
$$
After multiplication by $\epsilon$, eq.~(\ref{eq:uKdV}) can be written as



\be
k(k^2u'' + 3u^2 - Vu)' + \epsilon kF' = 0,   \label{eq:uF}
\ee


$$
kF' = \prt_tu + 6u\prt_xu + \frac{(\tilde k^3-k^3)}{\epsilon}u''' + \prt_t\theta_*u' + 6\prt_x\theta_*uu' + 3\tilde k^2\prt_xu''+\frac{3}{2}\prt_x(\tilde k^2)u'' +
$$

\be
+ \epsilon\left(3\tilde k\prt_{xx}u' + 3\prt_x\tilde k\prt_xu' + \prt_{xx}\tilde ku'\right) + \epsilon^2\prt_{xxx}u.   \label{eq:F'v}
\ee
As in section \ref{all1}, two exact secularity conditions are obtained from eq.~(\ref{eq:uF}) as $\overline{F'} = 0$ and $\overline{uF'} = 0$. Using the notation $Q = \overline{u}$, $Q_n = \overline{u^n}, n>1$, $G = k^2\overline{(u')^2}$, they read




\be
\prt_tQ + 3\prt_xQ_2 + \epsilon^2\prt_{xxx}Q = 0,   \label{eq:s1v}
\ee

\be
\prt_tQ_2 + 4\prt_xQ_3 - 3\prt_x\left(\frac{\tilde k^2}{k^2}G\right) + 3\epsilon\prt_x\left(\tilde k\overline{(u\prt_xu'-u'\prt_xu)}\right) + \epsilon^2\prt_x\left(\prt_{xx}Q_2 - 3\overline{(\prt_xu)^2}\right) = 0.  \label{eq:s2v}
\ee
To derive eq.~(\ref{eq:s2v}) we used the same identities as listed after eq.~(\ref{eq:S2}), only with $k$ replaced by $\tilde k$ there. Next we bring eqs.~(\ref{eq:s1v}), (\ref{eq:s2v}) to the form with fewer dependent variables, exactly like in section \ref{all1}.

\subsection{Transformation of the system (\ref{eq:s1v})-(\ref{eq:s2v}).}

We integrate eq.~(\ref{eq:uF}) to get

\be
k^2u'' + 3u^2 - Vu + \epsilon F = C_1,  \label{eq:I1v}
\ee

\ni where $C_1=C_1(x,t)$ is an arbitrary slow variable and $F$ is now defined as 

$$
kF = \prt_tJ_1 + 3\prt_xJ_2 + \frac{(\tilde k^3-k^3)}{\epsilon}u'' + \prt_t\theta_*u + 3\prt_x\theta_*u^2 + 3\tilde k^2\prt_xu'+\frac{3}{2}\prt_x(\tilde k^2)u' + 
$$

\be
+ \epsilon\left(3\tilde k\prt_{xx}u + 3\prt_x\tilde k\prt_xu + \prt_{xx}\tilde ku\right) + \epsilon^2\prt_{xxx}J_1,   \label{eq:Fv}
\ee
where functions $J_n$ are again defined by eqs.~(\ref{eq:hJn}) and (\ref{eq:avh}). Multiplying eq.~(\ref{eq:I1v}) by $2u'$ and integrating over $\theta$ one obtains (consistently with definitions (\ref{eq:hJn}) and (\ref{eq:avh}))

\be
k^2(u')^2 + 2u^3 - Vu^2 + 2\epsilon(uF - \int uF') = 2C_1u + C_2,   \label{eq:g1v}
\ee

\ni where $C_2$ is the arbitrary `integration slow variable' and $\int uF'$, as follows from eq.~(\ref{eq:F'v}), has form

$$
k\int uF' = \frac{\prt_tJ_2}{2} + \prt_x(2J_3 - \frac{3}{2}G_1) + \frac{(\tilde k^3-k^3)}{\epsilon}((uu')' - \frac{3}{2}(u')^2) + \prt_t\theta_*\frac{u^2}{2} + 2\prt_x\theta_*u^3 + 3\tilde ku\prt_x(\tilde ku') 
$$



\be
+ \epsilon\left( \frac{3}{2}\prt_x(\tilde k\int(u\prt_xu'-u'\prt_xu)) + \frac{u}{2}(3\tilde k\prt_{xx} + 3\prt_x\tilde k\prt_x + \prt_{xx}\tilde k)u \right) + \frac{\epsilon^2}{2}\left(\prt_{xxx}J_2 - 3\prt_x\int(\prt_xu)^2\right).   \label{eq:IuF'v}
\ee

\ni Here we defined $G_1 = \tilde k^2\int(u')^2 = \frac{\tilde k^2}{k^2}G\theta + \hat G_1$, $\overline{\hat G_1} = 0$, so that $\overline{G_1} = \tilde k^2/k^2\cdot G/2$ and the other formulas in eqs.~(\ref{eq:Jav}) and (\ref{eq:Mav}) of section \ref{all1} still hold.



\begin{prop}
$$
\prt_tQ + \prt_x(VQ+C_1) - \epsilon\prt_x\left(\frac{\prt_t\theta_*Q + \prt_x\theta_*(VQ+C_1)}{\tilde k}\right) -
$$

\be
- \epsilon^2\prt_x\left(2\prt_{xx}Q + \frac{3\prt_x\tilde k}{\tilde k}\prt_xQ + \frac{\prt_{xx}\tilde k}{\tilde k}Q\right) = 0,   \label{eq:S1KdV}
\ee

$$
\prt_tP + \prt_xH - \epsilon\left[ \prt_tQ_* + \prt_x(VQ_*) + \prt_x\left(\frac{\prt_t\theta_*P + \prt_x\theta_*H}{\tilde k}\right) \right] -
$$

$$
 - \epsilon^2\left[ \prt_tQ_d + \prt_x(VQ_d) - \prt_x\left(\frac{(\prt_t\theta_* + V\prt_x\theta_*)Q_*}{\tilde k}\right) + \prt_x\left(\frac{\prt_{xx}P}{2} + \frac{3\prt_x\tilde k}{2\tilde k}\prt_xP + \frac{\prt_{xx}\tilde k}{\tilde k}P\right) \right] +
$$

$$
+ \epsilon^3\prt_x\left[\frac{(\prt_t\theta_* + V\prt_x\theta_*)Q_d}{\tilde k} + \frac{\prt_{xx}Q_*}{2} + \frac{3\prt_x\tilde k}{2\tilde k}\prt_xQ_* + \frac{\prt_{xx}\tilde k}{\tilde k}Q_* \right] + 
$$

\be
+ \epsilon^4\prt_x\left(\frac{\prt_{xx}Q_d}{2} + \frac{3\prt_x\tilde k}{2\tilde k}\prt_xQ_d + \frac{\prt_{xx}\tilde k}{\tilde k}Q_d\right) = 0,  \label{eq:S2KdV}
\ee
where we denoted (recall that $\tilde k = k + \epsilon\prt_x\theta_*$) $P=VQ+C_1$, $H = VP - 3C_2$,

\be
Q_* = \frac{\prt_t\theta_*Q + \prt_x\theta_*(VQ+C_1)}{\tilde k},  \qquad  Q_d = \frac{3\prt_x(\tilde k\prt_xQ) + \prt_{xx}\tilde kQ}{\tilde k}.   \label{eq:PH*d}
\ee
\end{prop}
The proof is similar to that of Proposition 1 and we give it in appendix \ref{a:proof2}.

\par The secularity equations (\ref{eq:S1KdV}), (\ref{eq:S2KdV}) and the kinematic equation (\ref{eq:k}) comprise the exact nonperturbative Whitham-KdV equations in physical variables $V, C_1, C_2, k, Q$ and $\theta_*$. As earlier, the system of Whitham PDEs is not closed but is a convenient starting point to get the Whitham equations to any needed higher order in $\epsilon$.

\subsection{System of the Whitham-KdV equations to order $\epsilon$.}

Recall the consideration in the beginning of section \ref{pert}. We keep the same definition eq.~(\ref{eq:g10}) here and obtain the same equations (\ref{eq:u0})--(\ref{eq:s20}) leading to the KdV Riemann invariants $r_1\le r_2\le r_3$ as in eq.~(\ref{eq:rKdV}) and to expressions in eq.~(\ref{eq:ph-r}). To order $\epsilon$, secularity equations (\ref{eq:S1KdV}) and (\ref{eq:S2KdV}) are

\be
\prt_tQ + \prt_x(VQ+C_1 - \epsilon Q_*) = 0,  \label{eq:s1v1}
\ee

\be
\prt_t(VQ+C_1) + \prt_x\left(V(VQ+C_1) - 3C_2\right) - \epsilon\left[\prt_tQ_* + \prt_x\left(2VQ_* + \frac{C_1\prt_t\theta_* - 3C_2\prt_x\theta_*}{k}\right)\right] = 0, \label{eq:s2v1}
\ee
where now $Q_*$ is the leading order of $Q_*$ in the previous subsection,

\be
Q_* = \frac{\prt_t\theta_*Q_0 + \prt_x\theta_*(VQ_0+C_1)}{k}.   \label{eq:Q*}
\ee
We subtract the leading order equation (\ref{eq:s10}) from eq.~(\ref{eq:s1v1}) and similarly eq.~(\ref{eq:s20}) from eq.~(\ref{eq:s2v1}). The result is the linear homogeneous closed PDE system for the phase shift $\theta_*$ and $O(\epsilon)$ correction to the mean $Q_1$, $\epsilon Q_1=Q-Q_0$,

\be
\prt_tQ_1 + \prt_x(VQ_1 - Q_*) = 0,   \label{eq:thQ1}
\ee

\be
\prt_t(VQ_1) + \prt_x(V^2Q_1) - \prt_tQ_* - 2\prt_x(VQ_*) - \prt_x\left( \frac{C_1\prt_t\theta_* - 3C_2\prt_x\theta_*}{k} \right) = 0.   \label{eq:thQ2}
\ee
It is convenient to work with eq.~(\ref{eq:thQ1}) and the combination $(\ref{eq:thQ2})-V\cdot(\ref{eq:thQ1})$ of the above two equations, 

\be
(\prt_tV + V\prt_xV)Q_1 -\prt_tQ_* - Q_*\prt_xV - \prt_x\left(VQ_* + \frac{C_1\prt_t\theta_* - 3C_2\prt_x\theta_*}{k}\right) = 0,   \label{eq:thQ-}
\ee
as the final PDE system which we study below.

\subsection{Solution for the step initial condition.}

Consider again the step IC eq.~(\ref{eq:stepIC}) and recall the GP solution determined by eqs.~(\ref{eq:GP}) and (\ref{eq:phase0}). We try to solve eqs.~(\ref{eq:thQ1}) and (\ref{eq:thQ-}) for $\theta_*$ and $Q_1$ in this case. From step IC we expect the problem to be self-similar; this leads us to assume that $\theta_*=\theta_*(m)$. Then, by the GP formula, its space and time derivatives are

\be
\prt_x\theta_* = \frac{\theta_*'(m)}{tv_2'(m)},   \qquad   \prt_t\theta_* = -\frac{v_2\theta_*'(m)}{tv_2'(m)}.   \label{eq:dth*}
\ee
It is now convenient to change independent variables from $x,t$ to $m,t$; for the derivatives we get

\be
\prt_x = \frac{1}{tv_2'(m)}\prt_m,  \qquad  \prt_t \to \prt_t - \frac{v_2(m)}{tv_2'(m)}\prt_m,   \label{eq:dxt}
\ee
where in the last formula the $t$-derivative at constant $x$ is expressed in terms of $t$-derivative at constant $m$ and $m$-derivative. For the step IC, we have, according to the formulas (\ref{eq:rKdV}) and (\ref{eq:ph-r}),

$$
r_1=0, \quad r_2=m, \quad r_3=1, \qquad k^2 = \frac{1}{4K^2(m)},   \qquad Q_0 = 2\frac{E(m)}{K(m)}-1+m,
$$

\be
V = 2(1+m), \qquad C_1 = (1-m)^2,  \qquad C_2 = -2(1-m)(1-m^2),   \label{eq:param}
\ee

\ni i.e.~all of these quantities are functions of $m$ only. We get for $Q_*$ in eq.~(\ref{eq:Q*}),

\be
Q_* = \frac{[C_1 - (v_2-V)Q_0]\theta_*'(m)}{tkv_2'(m)} = \frac{q_*(m)}{t},  \qquad  q_*(m) = \frac{[C_1 - (v_2-V)Q_0]\theta_*'(m)}{kv_2'(m)}.   \label{eq:q*}
\ee
Thus, eqs.~(\ref{eq:thQ1}) and (\ref{eq:thQ-}) become (in the formulae below `prime' means taking the derivative with respect to $m$ of the corresponding function of $m$)

\be
\prt_tQ_1 - \frac{(v_2-V)}{tv_2'}\prt_mQ_1 + \frac{V'Q_1}{tv_2'} - \frac{q_*'}{t^2v_2'} = 0,   \label{eq:thq1}
\ee

\be
-\frac{(v_2-V)V'}{tv_2'}Q_1 + \frac{q_*}{t^2} + \frac{(v_2-V)}{t^2v_2'}q_*' - \frac{2V'}{t^2v_2'}q_* + \frac{1}{t^2v_2'}\left(\frac{(C_1v_2 + 3C_2)\theta_*'}{kv_2'}\right)' = 0.   \label{eq:thq-}
\ee
One is thus led to take $Q_1$ of the form $Q_1 = q_1(m)/t$, which is consistent with the implication of similarity i.e.~the $\epsilon$-expansion is in fact an expansion in powers of $\epsilon/t$. After using the last ansatz, the PDEs become ODEs in $m$, respectively,

\be
[(v_2-V)q_1 + q_*]' = 0,  \label{eq:q1d}
\ee

\be
-V'[(v_2-V)q_1 + q_*] + \left( (v_2-V)q_* + \frac{(C_1v_2 + 3C_2)\theta_*'}{kv_2'} \right)' = 0.   \label{eq:th*d}
\ee
Eq.~(\ref{eq:q1d}) integrates to give

\be
(v_2-V)q_1 + q_* = 3s_1 = \text{const.},   \label{eq:q1I}
\ee
and, after using this in eq.~(\ref{eq:th*d}), that equation becomes a total derivative and integrates to

\be
(v_2-V)q_* + \frac{(C_1v_2 + 3C_2)\theta_*'}{kv_2'} - 3s_1V = s_2 = \text{const.}   \label{eq:th*I}
\ee
Thus, recalling the definition of $q_*$ in eq.~(\ref{eq:q*}), we see that $\theta_*'(m)$ is determined from eq.~(\ref{eq:th*I}) and then $q_1(m)$ (and so $Q_1=q_1(m)/t$) is found from eq.~(\ref{eq:q1I}).   
\par We note that there are two unknown constants $s_1$ and $s_2$ in this equation. We find that  $\theta_*'(m)$ has a singularity as $m\to 1$. In order to make the singularity of $\theta_*'$ as $m\to1$ in eq.~(\ref{eq:th*I}) the mildest possible, one has to choose $s_2$ such that

\be
3s_1V+s_2 = 6s_1(1+m)+s_2 = -6s_1(1-m).   \label{eq:sth*}
\ee
Formulas (\ref{eq:param}), (\ref{eq:GP}), together with

\be
K'(m) = \frac{E(m)-(1-m)K(m)}{2m(1-m)},  \qquad  E'(m) = -\frac{K(m)-E(m)}{2m},   \label{eq:dKE}
\ee
yield

{\small $$
VC_1+3C_2 = -4(1+m)(1-m)^2,  \qquad  Q_0 = (1-m)\left(1 - \frac{8m}{v_2-V}\right), 
$$}

{\small $$
2C_1-(v_2-V)Q_0 = (1-m)\left(2(1 + 3m) - (v_2-V)\right),  \qquad v_2-V = -\frac{2K(m)}{K'(m)},
$$}

{\small $$
v_2'(m) = 2\left(1-(K/K')'\right) = \frac{2KK''}{(K')^2},   \qquad  K''(m) = \frac{2(2m-1)E(m) + (1-m)(2-3m)K(m)}{4m^2(1-m)^2}.
$$}
These expressions and eq.~(\ref{eq:sth*}), when substituted into eq.~(\ref{eq:th*I}), give the final formula for $\theta_*'$, with the still undetermined constant $s_1$,

{\small \be
\theta_*'(m) = \frac{3s_1[2(2m-1)E(m) + (1-m)(2-3m)K(m)]}{2(1-m)[(1+m)E^2(m) + 2(4m^2+m-1)K(m)E(m) + (1-m^2)(1-3m)K^2(m)]}.   \label{eq:th*'}
\ee}
Then one can verify that

$$
\theta_*'(m) \approx \frac{3s_1}{4(1-m)\ln(16/(1-m))} \quad \text{as } m\to1,
$$
but even this mildest possible singularity is not integrable, therefore we do not get nonsingular solution for $\theta_*$. Moreover, using an (equivalent) representation

$$
\theta_*' = \frac{3s_1K''}{2[K^2 + (1+3m)KK' + (1-m^2)(K')^2]},
$$
function $q_1(m)$ is found from eq.~(\ref{eq:q1I}) in the form

\be
q_1(m) = \frac{3s_1K'}{4K}\cdot\frac{[(1-m)(5m-1)(K')^2 - 8mKK' - 2K^2]}{[(1-m^2)(K')^2 + (1+3m)KK' + K^2]}.   \label{eq:q1}
\ee
As $m\to1$, $K(m)\approx \ln(16/(1-m))/2$ and $K'(m)\approx 1/(2(1-m))$, so one finds

$$
q_1(m) \approx -\frac{3s_1}{2(1-m)\ln(16/1-m)} \quad \text{as } m\to1,
$$
the same non-integrable singularity as $\theta_*'(m)$ has there. Recalling that $Q_1=q_1/t$ is the next-to-leading order correction to the mean $\overline u$ over the period, this last singularity looks particularly unrealistic. To remove it we need to take $s_1=0$ which would give $\theta_*= \text{const}$ and return us to the situation of sections \ref{all1} and \ref{pert}. 
\par In principle, one could ascribe the obtained singularities at the leading (solitonic) edge of the DSW region to the necessity of an intermediate/transition layer there with different scaling not captured by the current theory. Still then the only way to make the constant $s_1$ consistent with our numerics in section \ref{num} is to make it very small, not larger than of order $\epsilon$. This again effectively brings us back to the setting of sections \ref{all1} and \ref{pert}. 
\par We also note another difficulty for the current approach: the kinematic equation (\ref{eq:k}) and the secularity equations (\ref{eq:S1KdV}), (\ref{eq:S2KdV}), may not suffice to solve for all the variables $V, C_1, C_2, Q$ and $\theta_*$.

\section{Whitham theory: linear case}  \label{lin}

Let us compare the above with what Whitham theory for linearized KdV yields. The leading order solution of the linearized KdV equation,

\be
\prt_tu + \epsilon^2\prt_{xxx}u = 0,   \label{eq:lKdV}
\ee
which solves $k^2u_0''' - Vu_0'=0$, can be written as

\be
u_0 = c + A\cos\theta = c + A\cos\left(\frac{\theta_0}{\epsilon} + \theta_*\right),   \label{eq:lol}
\ee
with

\be
\prt_x\theta_0 = k,   \qquad  \prt_t\theta_0 = -kV,   \label{eq:ltheta}
\ee
$c$, $A$, $V$ and $\theta_*$ being independent parameters. The linear dispersion relation (or fixed constant period condition which is equivalent to it here) implies

\be
k = (-V)^{1/2},  \label{eq:lk2}
\ee
where we have in effect chosen a normalization for $\theta_0$ and the period. Then conservation of waves is a Hopf equation for $V$,

\be
\prt_tV + 3V\prt_xV = 0.   \label{eq:HopfV}
\ee
The first usual secularity condition for the next order equation yields $\prt_t\overline u_0=0$ which means that $c$ is a constant (not slowly varying). The second secularity condition from section \ref{all1},
\[ \int_0^1 [u_0(\partial_t u_0)+3k^2\partial_x u_0''+\frac{3}{2}\partial_x(k^2)u_0'']d\theta=0, \]
fixes the evolution equation for the amplitude $A$ as

\be
\prt_tA^2 + 3\prt_x(VA^2) = 0.   \label{eq:ampl}
\ee

\ni The correction $u_1$ has even and odd parts. The even part of the correction $u_{1e}$ is again tied with $\theta_*$ as 

\be
[k^3(u_{1e}'' + u_{1e}) + D\theta_*u_0 - 3cV\prt_x\theta_*]' = 0,   \label{eq:lu1e}
\ee
where $D\equiv \prt_t+3V\prt_x$. The general even solution of eq.~(\ref{eq:lu1e}) reads

\be
k^3u_{1g} = -\frac{AD\theta_*}{2}\theta\sin\theta - C_e + \gamma_0A\cos\theta,   \label{eq:lu1g} 
\ee
with arbitrary slow $C_e$ and $\gamma_0$. Enforcing periodicity yields the first order PDE for $\theta_*$,

\be
\prt_t\theta_* + 3V\prt_x\theta_* = 0.   \label{eq:lpde*}
\ee
Thus, here, in contrast to the nonlinear case of previous sections, the dynamics of $\theta_*$ is fixed by a usual secularity condition controlling periodicity rather than growth of the solution $u$ as function of fast variable $\theta_0/\epsilon$. Then the IC $u_{1e}(x,0)=0$ is achieved if one takes $C_e(x,0)=\gamma_0(x,0)=0$.
The solutions to both eq.~(\ref{eq:ampl}) and eq.~(\ref{eq:lpde*}) depend on the solution of Hopf equation (\ref{eq:HopfV}) whose general solution is implicitly given by

\be
x = 3Vt + x_0(V),  \qquad x_0(V(x,0)) =x,   \label{eq:Vl}
\ee
for a given initial condition $V(x,0)$. E.g.~if the IC for $\theta_*$ is written as $\theta_*(x,0)=\phi(V(x,0))$, then $\theta_*(x,t)$ is implicitly given by

\be
x = 3\phi^{-1}(\theta_*)t + x_0(\phi^{-1}(\theta_*)).   \label{eq:*sol}
\ee
These formulas can describe the evolution of the (total) phase given an initial condition for slow variables. E.g.~if $\theta_*(x,0)=\text{const.}$ initially, it remains constant for all times.
 
\par On the other hand, e.g.~for the step IC given by eq.~(\ref{eq:stepIC}), the exact solution of linearized KdV is obtained by Fourier transform (the analog of IST here); it is

\be
u(x,t) = \int_\xi^\infty \text{Ai}(\zeta)d\zeta,   \qquad  \xi=\frac{x}{(3\epsilon^2t)^{1/3}}.   \label{eq:uex}
\ee
Only its asymptotics for $\xi \ll -1$ can be described by periodic Whitham theory; it has indeed the form of leading order solution (\ref{eq:lol}),

\be
\xi \ll -1:  \qquad u(x,t) \approx 1 + \frac{1}{\sqrt\pi|\xi|^{3/4}}\cos\left(\frac{2}{3}|\xi|^{3/2}-\frac{3\pi}{4}\right).   \label{eq:uasympt}
\ee
This corresponds to the following solution for the Whitham variables: constant $c=1$ and 

\be
V = \frac{x}{3t},  \qquad A^2 = \frac{1}{\pi|\xi|^{3/2}} = \frac{\epsilon(3t)^{1/2}}{\pi|x|^{3/2}},   \qquad \frac{\theta_0}{\epsilon} = -\frac{2}{3}|\xi|^{3/2}=-\frac{2|x|^{3/2}}{3\epsilon(3t)^{1/2}},  \qquad \theta_* = \frac{3\pi}{4}.   \label{eq:Wl}
\ee
With this assignment all equations (\ref{eq:HopfV}), (\ref{eq:ampl}), (\ref{eq:ltheta}) with eq.~(\ref{eq:lk2}) and (\ref{eq:lpde*}) are satisfied. We see, however, that $V$ and $\theta_0$ are singular as $t\to0$; these solutions result from the asymptotics of the exact solution given in eq.~(\ref{eq:uex}). Any $\theta_*=\text{const}$ is a solution of eq.~(\ref{eq:lpde*}) but one needs to enforce ICs to get the correct value $3\pi/4$ from Whitham theory. It is unclear how to do this only within the context of Whitham theory due to the singular nature of $V$ and $\theta_0$. The numerics below clearly show that the phase shift is constant; there is only small ``numerical noise" around $\theta_* = \frac{3\pi}{4}$ in this linear case.

\subsection{Linear KdV Numerics}

\begin{figure} [ht]
\centering
\includegraphics[scale=.2]{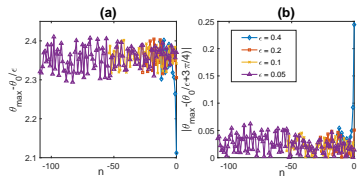}
\caption{Difference between $\theta_{\max}$ (see text) and (a) $\theta_0/\epsilon$; (b) $\theta_0/\epsilon + \theta_*$. Each curve corresponds to a different value of $\epsilon$.} 
\label{linear_compare_phase_shift}
\end{figure}

In the case of the linearized KdV equation, for $\xi \ll - 1$, the asymptotic solution is eq.~(\ref{eq:uasympt}) implying eq.~(\ref{eq:Wl}). The maxima of the function $\cos(\theta)$ occur when $\theta_{\max} = 2 \pi n$, $n \in \mathbb{Z}$. A comparison of the exact phase maxima and the asymptotic approximations given in eq.~(\ref{eq:uasympt}) is shown in Fig.~\ref{linear_compare_phase_shift}.

The difference between the exact maxima and the $O(1/\epsilon)$ approximation is found to yield a nearly constant value for all values of $n$. In Fig.~\ref{linear_compare_phase_shift}(b) the constant value is found to indeed be the phase shift $\theta_* = 3 \pi /4$ to within $O(\epsilon)$ error bounds. As $\epsilon$ decreases, $\xi$ decreases, resulting in a better asymptotic approximation. This improvement, as $\epsilon$ decreases, is different from the nonlinear case shown in Fig.~\ref{combined_phase_shift} where there is more structure and indications of possible intermediate/transition regions in the neighborhoods of $m=0$, $m=1$ as $\epsilon \to 0$. In Fig.~\ref{linear_compare_phase_shift} we clearly observe only some small random fluctuations around the known constant phase shift which do not imply any functional dependence. This difference with the nonlinear case is expected and lends additional support to our theoretical considerations. 

\section{Conclusion} 

We developed higher-order Whitham theory with a single fast phase for the KdV equation. This allowed us to determine the slow phase shift in the leading order solution and show that it is asymptotically constant for a wide range of initial conditions including step or steplike ICs. Other analytical possibilities which could be consistent with existing analytical results were ruled out. The role of nonlinearity and existence of a nontrivial $O(\epsilon)$ phase shift due to it predicted by our analysis 
is also clearly seen from comparison of our numerics for KdV. We also studied the linearized KdV equation and found that its phase shift is a constant. 

\par The conclusion that the phase shift as a function of space and time is asymptotically constant illuminates why for so many years since the seminal work~\cite{GurPit73} the leading order solution with constant phase shift is so widely and effectively used in applications; see e.g.~recent reviews~\cite{ElHoefRev16, ElHSh17} and their rather comprehensive lists of references. We note that considerable work has been devoted to the leading order solutions for multiple phases, see e.g.~\cite{ElHoefRev16, Grava17} and references therein.

\par Certain interesting and important issues remain for future work, in particular, the clarification of the relationship between small dispersion and long time limits originating from the space and time scaling properties of KdV equation. Numerical results also indicate the existence of intermediate/transition regions around the DSW edges. Their analytical description also presents an important problem for future work. To our knowledge, these transition regions from leading order DSW to constant solutions for steplike initial/boundary conditions have not been analytically described except for some partial results at leading order by IST in~\cite{AbBal13} and by matched asymptotic expansions assuming linear initial approximation and various matching conditions in several regions for the pure step problem in~\cite{LeaNee14}, both for the long-time regime.

\bigskip
{\bf\large Acknowledgement.} We thank the referees for comments which helped us substantially improve the text. This work was partially supported by the National Science Foundation under grant number DMS-1712793.


\def\thesection{Appendix~\Alph{section}:}


\appendix
\section{Non-secular corrections to the solution}  \label{a:corr1}

In general, formulas (\ref{eq:u1s}), (\ref{eq:u2s}) should be written as

\be
u_1 = \frac{u_0'(\theta)\int_{\theta(x,0)}^\theta w_2(z)F_1(z)dz - w_2(\theta)\int_{\theta(x,0)}^\theta u_0'(z)F_1(z)dz}{k^2W} + A_1u_0'(\theta) + A_2w_2(\theta),    \label{eq:u1a}
\ee

\be
u_2 = \frac{u_0'(\theta)\int_{\theta(x,0)}^\theta w_2(z)F_2(z)dz - w_2(\theta)\int_{\theta(x,0)}^\theta u_0'(z)F_2(z)dz}{k^2W} + A_{21}u_0'(\theta) + A_{22}w_2(\theta).    \label{eq:u2a}
\ee
Requiring periodicity of $u_1$ in eq.~(\ref{eq:u1a}) leads one to consider the difference $u_1(\theta+1)-u_1(\theta)$ (recall that we normalize $\theta$ to have period $1$). We have 

\be
w_2(\theta+1) - w_2(\theta) = K_2u_0'(\theta),   \qquad  K_2 = \alpha Q + \beta,    \label{eq:perw2}
\ee
see end of section \ref{pert}. Let $\theta_i=\theta(x,0)$. Consider the numerator of the first term in eq.~(\ref{eq:u1a}). We have, using periodicity of $u_0'$ and $F_1$ and eq.~(\ref{eq:perw2}),

$$
\Delta_1 \equiv u_0'(\theta+1)\int_{\theta_i}^{\theta+1} w_2(z)kF_1(z)dz - w_2(\theta+1)\int_{\theta_i}^{\theta+1} u_0'(z)kF_1(z)dz - 
$$

$$
-\left(u_0'(\theta)\int_{\theta_i}^\theta w_2(z)kF_1(z)dz - w_2(\theta)\int_{\theta_i}^\theta u_0'(z)kF_1(z)dz\right) = 
$$

\be
= u_0'(\theta)\int_{\theta}^{\theta+1} w_2(z)kF_1(z)dz - K_2u_0'(\theta)\int_{\theta_i}^{\theta+1} u_0'(z)kF_1(z)dz - w_2(\theta)\int_{\theta}^{\theta+1} u_0'(z)kF_1(z)dz.   \label{eq:perN1}
\ee
The antiderivative $\int u_0'(z)F_1(z)dz$ is an explicit odd periodic function found by direct integration which can be written as

\be
\Lambda_1 \equiv k\int u_0'(z)F_1(z)dz =       \label{eq:Iu0'F1}
\ee

$$
= (D + 2\prt_xV)\left(\frac{\hat u_0\hat J_1}{2} + \left(Q-\frac{V}{6}\right)\hat J_1 + \frac{k^2u_0'}{6}\right) + \frac{\hat u_0D\hat J_1 - \hat J_1D\hat u_0}{2} + \frac{k}{5}\prt_x\left(\frac{\alpha}{k}\hat J_1 + 2k\left(u_0-\frac{V}{6}\right)u_0'\right),   
$$
where we denoted $D = \prt_t + V\prt_x$, $\hat u_0 = u_0 - Q$, $\hat J_1 \equiv \hat J_1^0$. Thus, the averages over the period

\be
\overline{\Lambda_1(\theta)} = 0,  \qquad \int_{\theta}^{\theta+1} u_0'(z)F_1(z)dz = \int_0^1 u_0'(z)F_1(z)dz = 0.   \label{eq:u0'F1av}
\ee
Next, writing $w_2(\theta) = K_2u_0'(\theta)\theta + \tilde w_2(\theta)$, where $\tilde w_2(\theta)$ is an even periodic function, and taking into account that $F_1(\theta)$ is odd periodic, we get

\be
\int_{\theta}^{\theta+1} \tilde w_2(z)F_1(z)dz = 0.  \label{eq:tw2F1av}
\ee
Thus, since $\int_{\theta}^{\theta+1}\Lambda_1(z)dz=0$ due to $\Lambda_1(z)$ being odd periodic by eq.~(\ref{eq:Iu0'F1}),

$$
k\int_{\theta}^{\theta+1} w_2(z)F_1(z)dz = K_2\int_{\theta}^{\theta+1} zu_0'(z)kF_1(z)dz = K_2\left.z\Lambda_1(z)\right|_\theta^{\theta+1} - K_2\int_{\theta}^{\theta+1}\Lambda_1(z)dz =
$$

\be
= K_2\Lambda_1(\theta) - K_2\int_{\theta}^{\theta+1}\Lambda_1(z)dz = K_2\Lambda_1(\theta).   \label{eq:w2F1av}
\ee
Finally, use of eqs.~(\ref{eq:u0'F1av}) and (\ref{eq:w2F1av}) in eq.~(\ref{eq:perN1}) yields

\be
\Delta_1 = K_2u_0'(\theta)\Lambda_1(\theta) - K_2u_0'(\theta)\int_{\theta_i}^{\theta} u_0'(z)kF_1(z)dz = K_2u_0'(\theta)\Lambda_1(\theta_i).   \label{eq:pN1}
\ee
This depends on the initial conditions in general. For the case when $\theta_i\equiv\theta(x,0)=0$, also $\Delta_1=0$ and the first term in eq.~(\ref{eq:u1a}) is periodic in $\theta$. Since it is zero at $t=0$, we obtain eq.~(\ref{eq:u1s}) indeed (i.e.~the slow variables $A_1=A_2=0$ in this case). In general, however, periodicity of $u_1$ in eq.~(\ref{eq:u1a}) is achieved by taking

\be
A_2 = -\frac{\Lambda_1(\theta_i)}{k^3W},  \label{eq:A2}
\ee
as is seen from eqs.~(\ref{eq:pN1}) and (\ref{eq:perw2}). Then the initial condition $u_1(x,0)=0$ is ensured by taking $A_1$ such that

\be
A_1u_0'(x,0) + A_2w_2(x,0) = 0.  \label{eq:A1}
\ee
Quite similar considerations apply to $u_2$ in eq.~(\ref{eq:u2a}) and further higher order corrections to the solution.

\section{Proof of Proposition 2}  \label{a:proof2}

Exactly as in section \ref{all1}, one obtains eqs.~(\ref{eq:I1av}), (\ref{eq:g1av}), (\ref{eq:uI1av}) and (\ref{eq:*2}). Taking into account the secularity condition eq.~(\ref{eq:s1v}) in eq.~(\ref{eq:Fv}) brings the ``forcing" $F$ to the explicitly periodic form,







$$
kF = \prt_t\hat J_1 + 3\prt_x\hat J_2 + \frac{(\tilde k^3-k^3)}{\epsilon}u'' + \prt_t\theta_*u + 3\prt_x\theta_*u^2 + 3\tilde k^2\prt_xu'+\frac{1}{2}\prt_x(\tilde k)^2u' + 
$$

\be
+ \epsilon\left(3\tilde k\prt_{xx}u + 3\prt_x\tilde k\prt_xu + \prt_{xx}\tilde ku\right) + \epsilon^2\prt_{xxx}\hat J_1.   \label{eq:Fpv}
\ee

\ni The last equation integrated over the period becomes 

\be
k\overline F = \prt_t\theta_*Q + 3\prt_x\theta_*Q_2 + \epsilon(3\tilde k\prt_{xx} + 3\prt_x\tilde k\prt_x + \prt_{xx}\tilde k)Q.   \label{eq:Favv}
\ee

\ni In turn, taking into account the secularity condition eq.~(\ref{eq:s2v}) lets one bring the quantity $\int uF'$ to explicitly periodic form and its average over the period reads:



 

$$
k\overline{\int uF'} = \prt_t\theta_*\frac{Q_2}{2} + 2\prt_x\theta_*Q_3 - \frac{3}{2}\frac{(\tilde k^3-k^3)}{\epsilon k^2}G +
$$

\be
+ \frac{3\tilde k^2}{2}\overline{(u\prt_xu'-u'\prt_xu)} + \epsilon\left[\frac{3\tilde k}{2}\left(\frac{\prt_{xx}Q_2}{2} - \overline{(\prt_xu)^2}\right) + \frac{3}{2}\prt_x\tilde k\frac{\prt_{x}Q_2}{2} + \prt_{xx}\tilde k\frac{Q_2}{2}\right].   \label{eq:IuF'avv}
\ee
Upon using eq.~(\ref{eq:IuF'avv}), eq.~(\ref{eq:*2}) acquires the form containing exactly the combination entering the secularity equation (\ref{eq:s2v}),


$$
4Q_3 - 3\frac{\tilde k^2}{k^2}G + 3\epsilon\tilde k\overline{(u\prt_xu'-u'\prt_xu)} - 3\epsilon^2\overline{(\prt_xu)^2} = 
$$

\be
= \frac{k}{\tilde k}(VQ_2-C_2) - \epsilon\frac{\prt_t\theta_*Q_2}{\tilde k} - \frac{\epsilon^2(3\prt_x(\tilde k\prt_xQ_2) + 2\prt_{xx}\tilde kQ_2)}{2\tilde k}.   \label{eq:*2v2}   
\ee
We substitute the right-hand side of eq.~(\ref{eq:*2v2}) into secularity condition eq.~(\ref{eq:s2v}) and the last becomes

$$
\prt_tQ_2 + \prt_x(VQ_2-C_2) - \epsilon\prt_x\left(\frac{\prt_t\theta_*Q_2 + \prt_x\theta_*(VQ_2-C_2)}{\tilde k} \right) -
$$

\be
 - \epsilon^2\prt_x\left(\frac{\prt_{xx}Q_2}{2} + \frac{3\prt_x\tilde k}{2\tilde k}\prt_xQ_2 + \frac{\prt_{xx}\tilde k}{\tilde k}Q_2\right) = 0.   \label{eq:s22v}
\ee

\ni Finally we use eqs.~(\ref{eq:I1av}) and (\ref{eq:Favv}) to express $Q_2$ as

\be
3Q_2 = VQ+C_1 - \epsilon\cdot \frac{\prt_t\theta_*Q + \prt_x\theta_*(VQ+C_1)}{\tilde k} - \epsilon^2\left( \frac{3\prt_x(\tilde k\prt_xQ)}{\tilde k} + \frac{\prt_{xx}\tilde k}{\tilde k}Q \right)   \label{eq:Q2v}
\ee
and substitute it into eqs.~(\ref{eq:s1v}) and (\ref{eq:s22v}). Thus, we obtain the secularity conditions in their final form, eqs.~(\ref{eq:S1KdV}) and (\ref{eq:S2KdV}), as claimed.


\section{KdV with step IC phases from IST/RHP} \label{a:EGT}


For the KdV equation with steplike ICs, the total phase including the phase shift was computed recently~\cite{EGKT13, EGT16} by solving a vector Riemann-Hilbert problem (RHP) and using the steepest descent approach of Deift and Zhou~\cite{DZ93, DVZ94}. The solution is constructed for long-time asymptotics rather than for the small dispersion limit. However, for the pure step, the former appear to be equivalent to the latter since, by rescaling $x$ and $t$, the KdV equation is seen to depend only on $x/\epsilon$ and $t/\epsilon$ while the IC (\ref{eq:stepIC}) does not depend on scaling. Reintroducing $\epsilon$ into the formulas of~\cite{EGKT13, EGT16} (where $\epsilon=1$) and taking the initial jump $c^2=1$ there, their result for the total phase in the DSW region reads

\be
\theta = \frac{tB(\xi)}{2\pi\epsilon} + \frac{\Delta(\xi)}{2\pi} \pm \frac{1}{2},   \qquad  \xi \equiv \frac{x}{12t},    \label{eq:Tphase}
\ee
where the leading order phase function is

\be
B(\xi) = 24\int_{\sqrt{m(\xi)}}^1\left(\xi + \frac{1-m(\xi)}{2} - s^2\right)\sqrt\frac{s^2-m(\xi)}{1-s^2}ds   \label{eq:Bfunc}
\ee
and the phase shift is determined by

\be
\Delta(\xi) = \frac{1}{K(m)}\int_{\sqrt{m(\xi)}}^1\frac{\log(4s\sqrt{1-s^2})}{\sqrt{(1-s^2)(s^2-m(\xi))}}ds.   \label{eq:Dfunc}
\ee
Here the elliptic modulus parameter $m(\xi)$ is implicitly given by the equation

\be
\int_0^m\left(\xi + \frac{1-m(\xi)}{2} - s^2\right)\sqrt\frac{m-s^2}{1-s^2}ds  = 0, \label{eq:TGP}
\ee
which is an equivalent form of Gurevich-Pitaevskii (GP) equation eq.~(\ref{eq:GP}). Note that in~\cite{EGKT13, EGT16} the solution is expressed in terms of a second log-derivative of elliptic theta-function; to match its total phase with that of the cnoidal function one has to add $\pm 1/2$ to their phase $\frac{tB(\xi)}{2\pi\epsilon} + \frac{\Delta(\xi)}{2\pi}$, see e.g.~\cite{BF71}. One can express the three last equations explicitly in terms of complete elliptic integrals. Using the identities, see e.g.~\cite{BF71},


{\small \be
\int_0^1\sqrt\frac{1-s^2}{1-ms^2}ds = \frac{E(m)-(1-m)K(m)}{m},   \qquad \int_0^1\sqrt{(1-s^2)(1-ms^2)}ds = \frac{(1+m)E(m)-(1-m)K(m)}{3m},   \label{eq:IJ}
\ee}

$$
I_0=\int_{\sqrt m}^1\frac{ds}{\sqrt{(1-s^2)(s^2-m)}} = K(1-m),  \qquad I_2=\int_{\sqrt m}^1\frac{s^2ds}{\sqrt{(1-s^2)(s^2-m)}} = E(1-m), 
$$

\be
I_4=\int_{\sqrt m}^1\frac{s^4ds}{\sqrt{(1-s^2)(s^2-m)}} = \frac{2(1+m)E(1-m)-mK(1-m)}{3},   \label{eq:I2n}
\ee

\ni one gets the GP equation eq.~(\ref{eq:GP}) from eq.~(\ref{eq:TGP}) and the expression for $B(\xi)$ in the form

{\small $$
\frac{B(\xi)}{24} = \left(\xi + \frac{1+m}{2}\right)I_2 - m\left(\xi + \frac{1-m}{2}\right)I_0 - I_4 =  \left(\frac{2(1+m)}{3} + \frac{m(1-m)K(m)}{3((1-m)K(m)-E(m))}\right)E(1-m) - 
$$}


{\small $$
- m\left(\frac{2-m}{3} + \frac{m(1-m)K(m)}{3((1-m)K(m)-E(m))}\right)K(1-m) - \frac{2(1+m)E(1-m)-mK(1-m)}{3}.
$$}
This is further simplified using the Legendre relation

$$
K(m)E(1-m) + K(1-m)E(m) - K(m)K(1-m) = \frac{\pi}{2},
$$
to get

$$
\frac{tB(\xi)}{2\pi\epsilon} = -\frac{t}{\epsilon}\cdot\frac{2m(1-m)}{E(m) - (1-m)K(m)},
$$
which exactly matches the fast phase eq.~(\ref{eq:phase0}) from Whitham theory. 
Also, using the formulas~\cite{BF71}, p.~288,

$$
\int_{\sqrt m}^1\frac{\log(s)ds}{\sqrt{(1-s^2)(s^2-m)}} = \frac{K(1-m)\ln m}{4},    
$$

\be
\int_0^{\sqrt{1-m}}\frac{\log(s)ds}{\sqrt{(1-s^2)(1-m-s^2)}} = \frac{K(1-m)\ln(1-m)}{4} - \frac{\pi K(m)}{4},   \label{eq:lnBF}
\ee
one brings the slow phase shift to the simple form

\be
\theta_*=\frac{\Delta(m(\xi))}{2\pi} \pm \frac{1}{2} = \frac{K(1-m)}{2\pi K(m)}\log[4(m(1-m))^{1/4}] - \frac{1}{8} \pm \frac{1}{2}.   \label{eq:phase1}
\ee
As $m\to1$, the phase shift approaches a constant

$$
m\to1:  \qquad \frac{\Delta}{2\pi} \pm \frac{1}{2} \to -\frac{1}{4} \pm \frac{1}{2}.      
$$
Note this is not just the $\pm \frac{1}{2}$ as it should be at this edge. In the other limit $m\to0$, $\Delta$ diverges as

$$
m\to0:  \qquad \frac{\Delta}{2\pi} \approx -\frac{(\ln m)^2}{8\pi^2}.   
$$
The last formulas imply that proper matching of the solutions in the DSW region and, respectively, the region ahead and the linear oscillatory region behind it should remove the discrepancy and the singularity.

\par We observe complete agreement for the {\it leading order} fast phase. The nontrivial result of~\cite{EGT16} for the next order phase shift contrasts sharply with our analytical and numerical findings. The reason is a matter for future investigation. A possible source of the discrepancy could be short times $t\sim\epsilon$ where the apparent equivalence of small dispersion and long time results for scale-invariant ICs could be broken by subtle effects related to interchange of the two limits. Also the above limits as $m\to1$ and $m\to0$ of the phase shift of~\cite{EGT16} at the very least imply the necessity of intermediate/transition regions at the edges of DSW. Their existence at both edges could be consistent with our numerical results in section \ref{num}. For decaying ICs, such regions have been described analytically in~\cite{ClGrPII, ClGrSol} in the small dispersion limit.




\begin{thebibliography}{1}  

\bibitem{Ab2011}
M.J.~Ablowitz.
Nonlinear dispersive waves: asymptotic analysis and solitons.
{\em Cambridge Univ. Press}, Cambridge, UK, 2011.


\bibitem{AbBal13}
M.J.~Ablowitz, D.~Baldwin.
Dispersive shock wave interactions and asymptotics.
{\em Phys. Rev. E}, 87:022906, 2013.

\bibitem{AbBe70}
M.J.~Ablowitz, D.~J.~Benney.
The evolution of multi-phase modes for nonlinear dispersive waves.
{\em Stud. in Appl. Math.}, 49:225--238, 1970.




\bibitem{AbCl91}
M.J.~Ablowitz, P.~Clarkson.
Solitons, nonlinear evolution equations and the inverse scattering.
{\em Cambridge University Press}, NY, 1991.


\bibitem{AbSeg81}
M.J.~Ablowitz, H.~Segur.
Solitons and the inverse scattering transform.
{\em SIAM}, Philadelphia, 1981.


\bibitem{BF71}
P.~Byrd, M.~Friedman.
Handbook of elliptic integrals for engineers and scientists.
{\em Springer-Verlag}, 2nd edition, Berlin, 1971.

\bibitem{ClGrPII}
T.~Claeys, T.~Grava.
Painlev\'e II asymptotics near the leading edge of the oscillatory zone for the Korteweg-de Vries equation in the small dispersion limit.
{\em Comm. Pure Appl. Math.}, LXIII:203--232, 2010.

\bibitem{ClGrSol}
T.~Claeys, T.~Grava.
Solitonic asymptotics for the Korteweg-de Vries equation in the small dispersion limit.
{\em SIAM J. Math. Anal.}, 42:2132--2154, 2010.



\bibitem{DVZ94}
P.~Deift, S.~Venakides, X.~Zhou.
The collisionless shock region for the long-time behavior of solutions of the KdV equation.
{\em Comm. Pure Appl. Math.}, XLVII:199--206, 1994.

\bibitem{DVZ97}
P.~Deift, S.~Venakides, X.~Zhou.
New results in small dispersion KdV by an extension of the steepest descent method for Riemann-Hilbert problems.
{\em Intern. Math. Res. Not.}, 1997:285--299, 1997.

\bibitem{DZ93}
P.~Deift, X.~Zhou.
A steepest descent method for oscillatory Riemann-Hilbert problems. Asymptotics for the MKdV equation.
{\em Ann. Math.}, 137:295--303, 1993.

\bibitem{EGKT13}
I.~Egorova, Z.~Gladka, V.~Kotlyarov, G.~Teschl.
Long-time asymptotics for the Korteweg-de Vries equation with steplike initial data.
{\em Nonlinearity}, 26:1839--1864, 2013.

\bibitem{EGT16}
I.~Egorova, Z.~Gladka, G.~Teschl.
On the form of dispersive shock waves of the Korteweg-de Vries equation.
{\em Journ. Math. Phys. Anal. Geom.}, 12:3--16, 2016.


\bibitem{GurPit73}
A.~Gurevich, L.~Pitaevskii.
Nonstationary structure of a collisionless shock wave.
{\em J. Exp. Theor. Phys.}, 38:291--297, 1974.

\bibitem{ElHoefRev16}
G.~El, M.~Hoefer.
Dispersive shock waves and modulation theory.
{\em Phys. D}, 333:11--65, 2016.

\bibitem{ElHSh17}
G.~El, M.~Hoefer, M.~Shearer.
Dispersive and diffusive-dispersive shock waves for nonconvex conservation laws.
{\em SIAM Review}, 59:3--61, 2017.

\bibitem{Grava17}
T.~Grava.
Whitham modulation equations and application to small dispersion asymptotics and long time asymptotics of nonlinear dispersive equations.
{\em Lect. Notes Phys. 926, Springer}, 2016; {\em arXiv:1701.00069}, 2017.

\bibitem{GraKle07}
T.~Grava, C.~Klein.
Numerical solution of the small dispersion limit of Korteweg-de Vries and Whitham equations.
{\em Comm. Pure Appl. Math.}, LX:1623--1664, 2007.

\bibitem{Hab88}
R.~Haberman.
The modulated phase shift for weakly dissipated nonlinear oscillatory waves of the Korteweg-de Vries type.
{\em Stud. Appl. Math.}, 78:73--90, 1988.



\bibitem{trefethen} 
A.-K.~Kassam, L.~N.~Trefethen.
Fourth-order time-stepping for stiff PDEs. 
{\em SIAM J. Sci. Comput.}, 26:1214--1233, 2005.



\bibitem{LLV93}
P.~Lax, C.~D.~Levermore, S.~Venakides.
The generation and propagation of oscillations in dispersive initial value problems and their limiting behavior.
{\em Important developments in soliton theory, A.~Fokas and V.~E.~Zakharov (eds.), Springer, Berlin,} 205--241, 1993.

\bibitem{LeaNee14}
J.~Leach, D.~Needham.
\newblock The large-time development of the solution to an initial-value problem for the Korteweg-de Vries equation: II. Initial data has a discontinuous compressive step.
\newblock {\em Mathematika}, 60:391--414, 2014.

\bibitem{Ven90}
S.~Venakides.
The Korteweg-de Vries equation with small dispersion: higher order Lax-Levermore theory.
{\em Comm. Pure Appl. Math.}, XLIII:335--361, 1990.


\bibitem{Whitham65}
G.~Whitham.
Nonlinear dispersive waves.
{\em Proc. Roy. Soc.}, 283:238--261, 1965.

\bibitem{Whitham74}
G.~Whitham.
Linear and nonlinear waves.
{\em Wiley}, NY, 1974.


\end{thebibliography}
\end{document}